\documentclass[12pt]{article}
\pdfoutput=1
\usepackage{amsmath}
\usepackage{graphicx,psfrag,epsf}
\usepackage{enumerate}
\usepackage{natbib}
\usepackage[T1]{fontenc}
\usepackage[latin9]{inputenc}
\usepackage{color}
\usepackage{float}
\usepackage{hyperref}
\usepackage{multirow}

\makeatletter

\providecommand{\tabularnewline}{\\}
\newcommand{\red}[1]{{\color{red} #1}} 

\makeatother

\begin{document}

\def\spacingset#1{\renewcommand{\baselinestretch}%
{#1}\small\normalsize} \spacingset{1}

\title{\bf  Enabling Interactivity on Displays of Multivariate Time Series and Longitudinal Data}
\author{Xiaoyue Cheng, Dianne Cook, Heike Hofmann \\
        Department of Statistics, Iowa State University}
\date{}
\maketitle

\bigskip
\begin{abstract}
Temporal data is information measured in the context of time. This contextual structure provides components that need to be explored to understand the data and that can form the basis of interactions applied to the plots. In multivariate time series we expect to see temporal dependence, long term and seasonal trends and cross-correlations. In longitudinal data we also expect within and between subject dependence. Time series and longitudinal data, although analyzed differently, are often plotted using similar displays.  We provide a taxonomy of interactions on plots that can enable exploring temporal components of these data types, and describe how to build these interactions using data transformations. Because temporal data is often accompanied other types of data we also describe how to link the temporal plots with other displays of data. The ideas are conceptualized into a data pipeline for temporal data, and implemented into the R package \texttt{\textbf{cranvas}}. This package provides many different types of interactive graphics that can be used together to explore data or diagnose a model fit. 

\end{abstract}

\noindent%


\section{Introduction}

Constructing interactive graphics for temporal data can be enabled by building upon static displays. Aspects of the graphical elements in the displays can be made accessible to modification by user actions, for the purpose of facilitating different exploration of the temporal components in the data. To explain how to do this we first need to understand how time might be structured, and the common types of temporal data displays. 

\subsection{Characterizing time}

In data, time is coded in many different ways: as a date/time format ("Wed Oct 15 09:51:53 2014"), as discrete or continuous values, sporadic events or intervals. Recoding a time variable into other units like week in the year or days in the month, although convenient for some tasks, brings imprecision, as do events like leap years, and seconds. There may not be a well-defined absolute time line, and periodicity can be hard to quantify. Variables measured over time may be measured on different scales, e.g. atmospheric particulate matter and mortality might be drawn from different sources to study the effect of pollution on human health and measured at different resolutions. 

The most common description of time is as a continuous or discrete ordered numerical variable. All data is technically discrete, but if measurements are recorded often enough, and long enough they are effectively continuous. For example, currency exchange rates change on a microsecond basis, blood pressure measurements made with a wearable device records at every minute. For these examples time could essentially be considered continuous. However, it may be not helpful to evaluate trends on this microscale, and aggregating at an hourly, daily or monthly value may be sufficient. For simplicity, the methods developed in this paper assume the time variable is measured on a discrete scale.

\begin{figure}[h]
\begin{centering}
\includegraphics[width=0.48\textwidth]{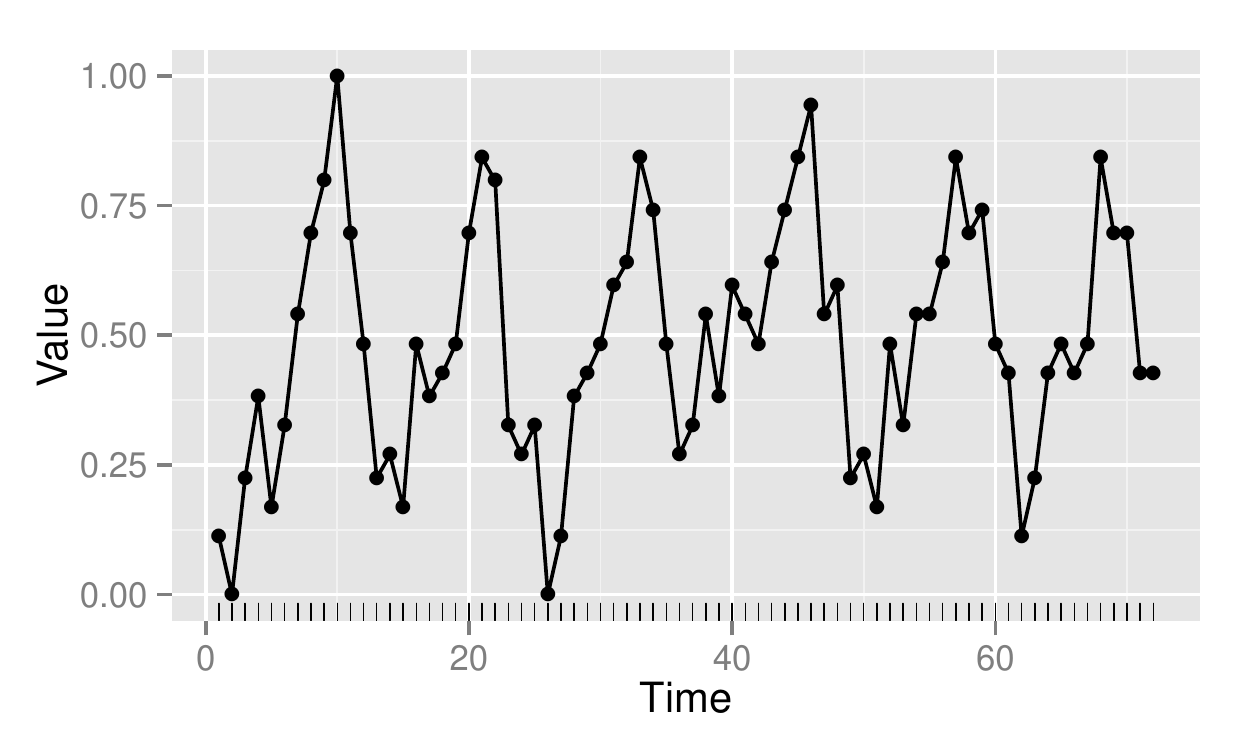}
\includegraphics[width=0.48\textwidth]{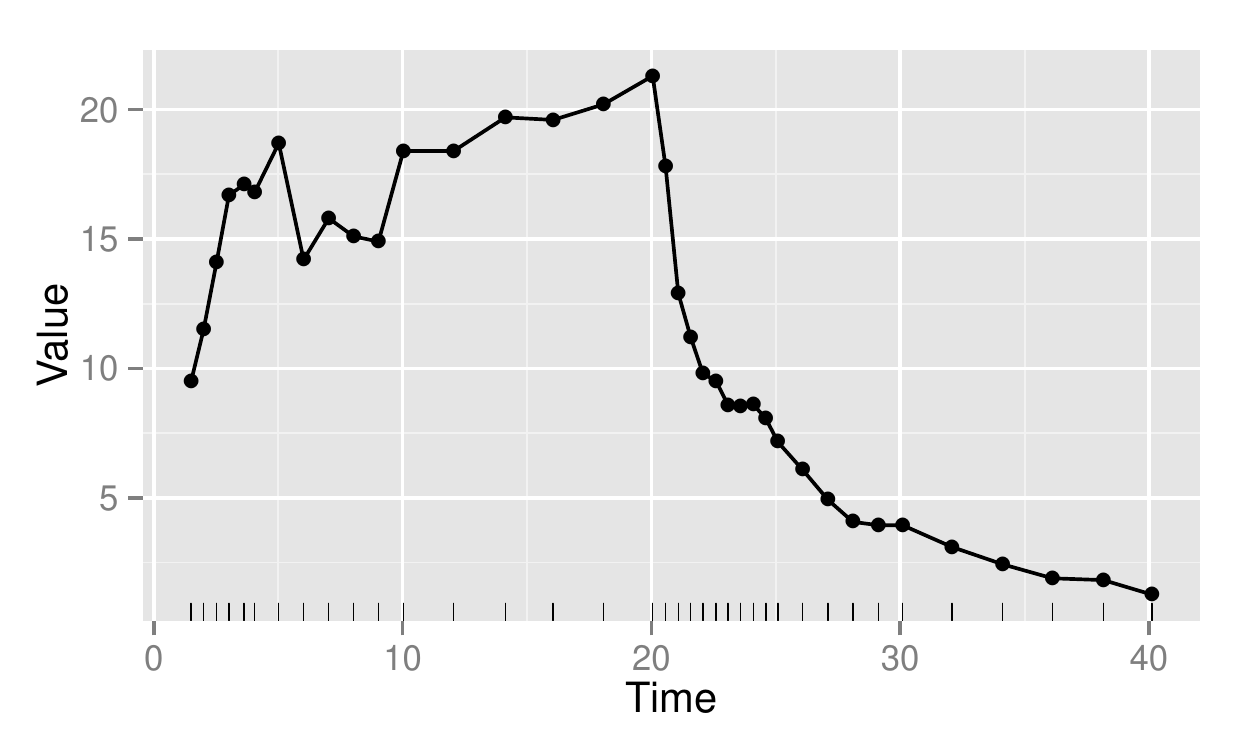}
\end{centering}
\caption{\label{fig:Time-series-plots}Time series plots for regular time spacing
 (left) and irregular time spacing  (right). Tick marks at bottom indicate the time sampling.}
\end{figure}

On a discrete scale, it may be possible to have regular or irregular time spacing. Regular time spacing means that the measurement is collected on constant time intervals, e.g. average monthly temperature in climate records. Irregular time spacing typically arise from events like measurements taken during visits to the doctor's office. Figure \ref{fig:Time-series-plots} illustrates data collected at regular, and irregular, intervals, respectively.

When many measurements are made more complications ensue. In climate records we may have temperature measurements taken at many different locations. In longitudinal data, we may have records for many patients. Making comparisons between many time series is a challenge. This work addresses this for a moderate number of series.

\subsection{Visualizing time}

Longitudinal data and time series data, although analyzed very differently, have in common the context of time that is commonly plotted in similar ways. Here are examples of common types of temporal displays.

Time is most conventionally displayed on a horizontal axis of a plot. There are many different variations: 

\begin{itemize} \itemsep 0in
\item A line graph, the basic building block for temporal data, displays the measured variable on the vertical axis, time horizontally, and consecutive time points are connected with line segments. Figure \ref{fig:Time-series-plots} gives two examples of line graphs, for a single measured variable: (left) a classical time series plot, (right) a profile plot of longitudinal data. When there are many series, for example time series of different stocks or different geographic locations, or many patients, the series may be overlaid on the same plot (Figure \ref{fig:horizontal-axis} row 1 left), or faceted in several blocks (Figure \ref{fig:horizontal-axis} row 1 right). 

\item Small multiples are used to display multiple series in separate plots (Figure \ref{fig:horizontal-axis} row 2 right). The terminology small multiples was introduced by \citet{tufte1983visual}. A special modification was developed for multiple times series, called sparklines \citet{tufte2006evidence}. Small multiples can also be generated by subsetting based on categorical covariates (e.g. \citet{cleveland1993}). 

\item Stacked graphs. Originated by \citeauthor{playfair2005playfair} in
1700's and recently discussed by \citet{byron2008stacked,javed2010graphical,heer2010tour},
a stacked graph draws the time series sequentially, and uses the previous
time series as the baseline for the current series (Figure \ref{fig:horizontal-axis} row 2 left). It is mostly used
for the longitudinal data rather than multivariate time series since
the individuals from the longitudinal data share the same scale.

\item Themeriver and streamgraph. Themeriver is created by \citet{havre2000themeriver},
which is a special case of the stacked graphs, since it moves the
starting baseline from the bottom to the center, and makes the plot
symmetric vertically (Figure \ref{fig:horizontal-axis} row 3 left). Streamgraph is developed later by \citet{byron2008stacked}.
It changed the algorithm to avoid the symmetry which increases the
internal distortion.

\item Horizon graphs. The horizon graph is inspired by two-tone pseudo coloring
\citep{saito2005two} and formally developed at Panopticon Software
\citep{reijner2008development}. Two-tone pseudo coloring is a technique
to visualize the details of multiple time series precisely and effectively.
However, the horizon graph became more popular after mirroring the
lower part of the series and simplifying the color scheme (Figure \ref{fig:horizontal-axis} row 3 right). The horizon
graphs were designed for visualizing the stock prices and economic/financial
data, so the features fit the requirements very well:  (1) The data
have a baseline, which is usually the value at the starting time point.
Then the baseline can be used to mirror the negative part to the positive,
in order to save the graph space, where `negative/positive' means
smaller/greater than the baseline.  (2) The positive and negative performance
should be distinguished, so the horizon graph provides two hues.  (3)
The number of the color bands should be small, usually three color
bands for the positive values and three for the negative. Finding
the band height is easy for the stock prices since they can use 10\%
of the initial value, and in most cases the price will not increase
or decrease for more than 30\%.

\begin{center}
\begin{figure}[htp]
\begin{centering}
\begin{tabular}{cc}
(a) & (b) \\
\includegraphics[width=0.48\textwidth]{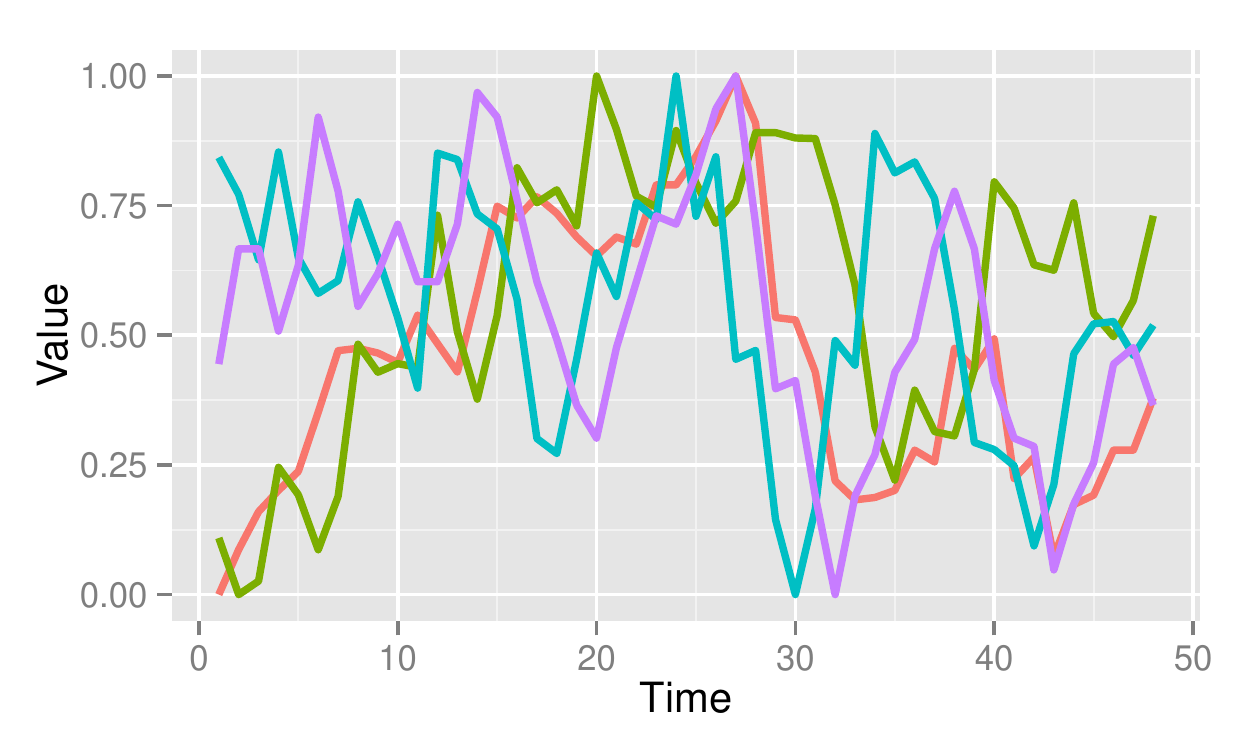} &
\includegraphics[width=0.48\textwidth]{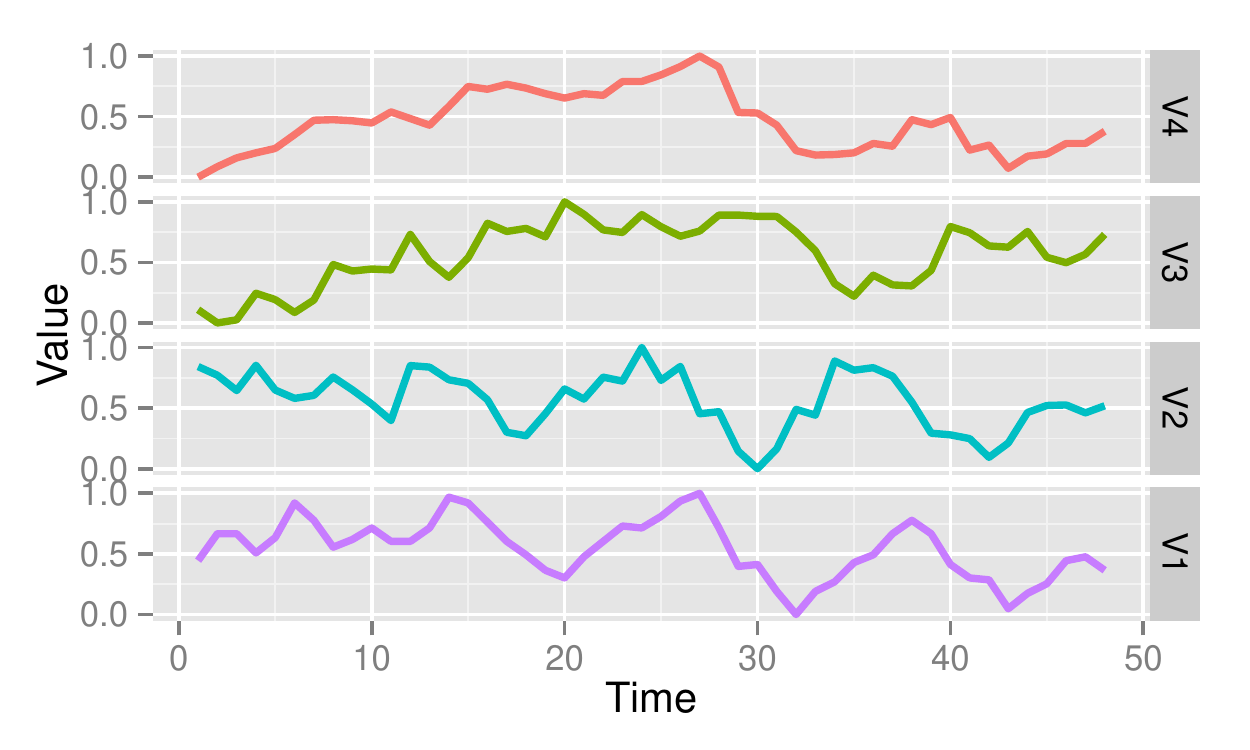}\\
(c) & (d) \\
\includegraphics[width=0.48\textwidth]{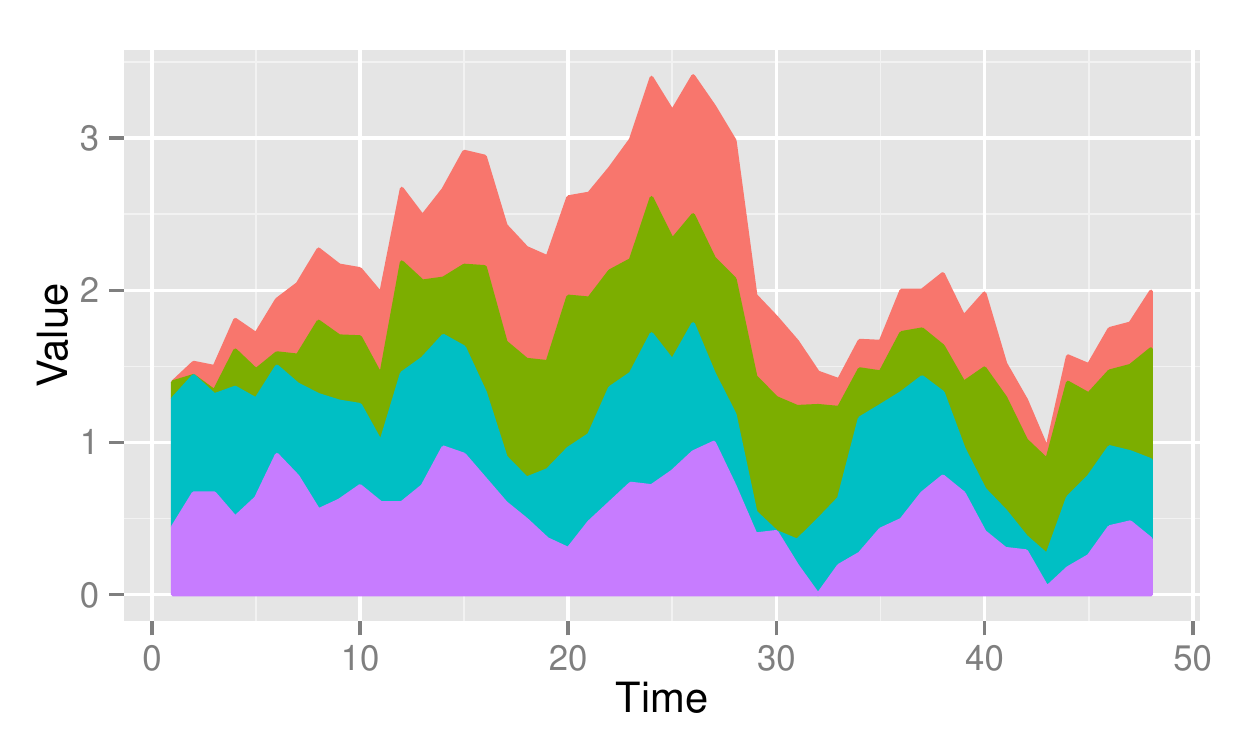} &
\includegraphics[width=0.48\textwidth]{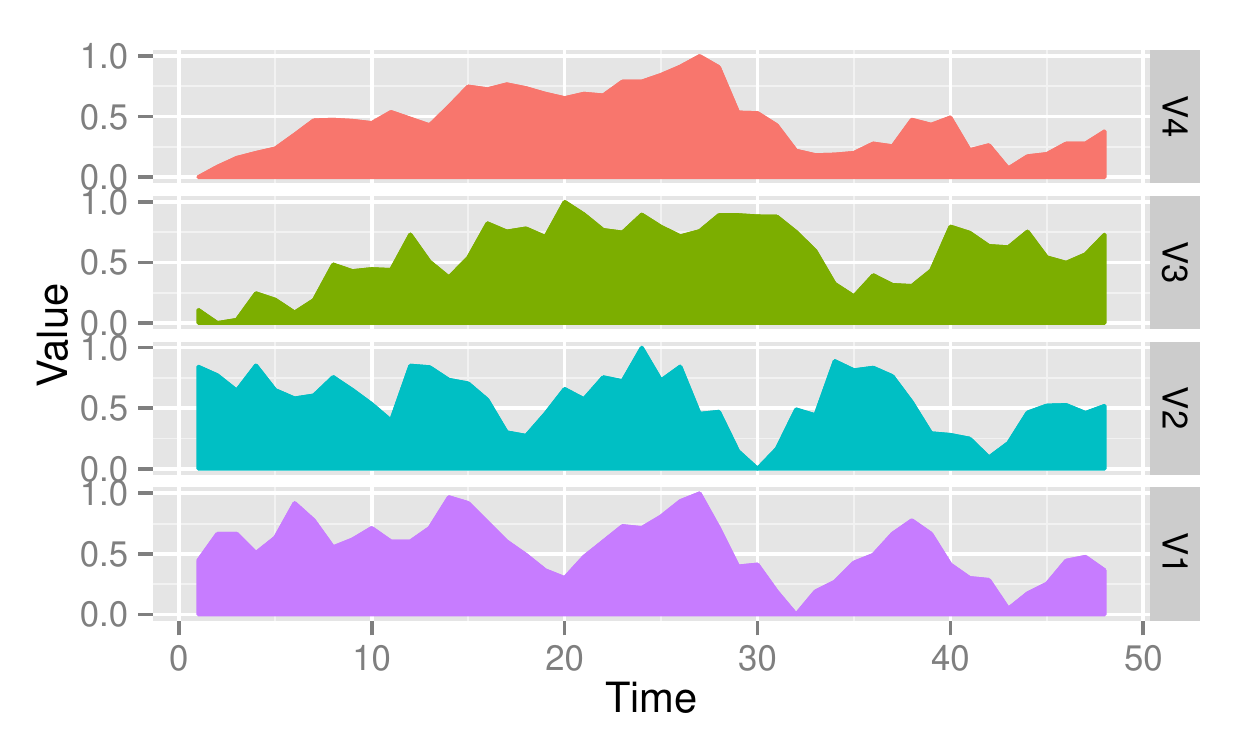}\\
(e) & (f) \\
\includegraphics[width=0.48\textwidth]{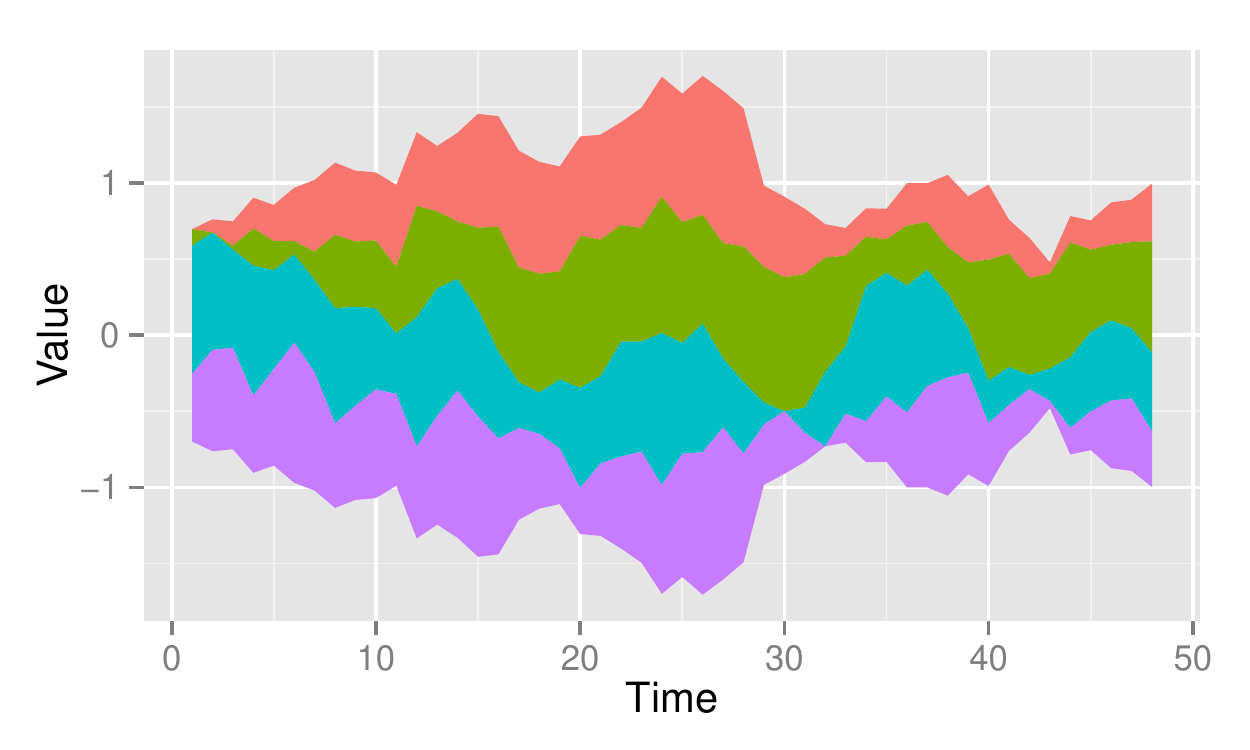} &
\includegraphics[width=0.48\textwidth]{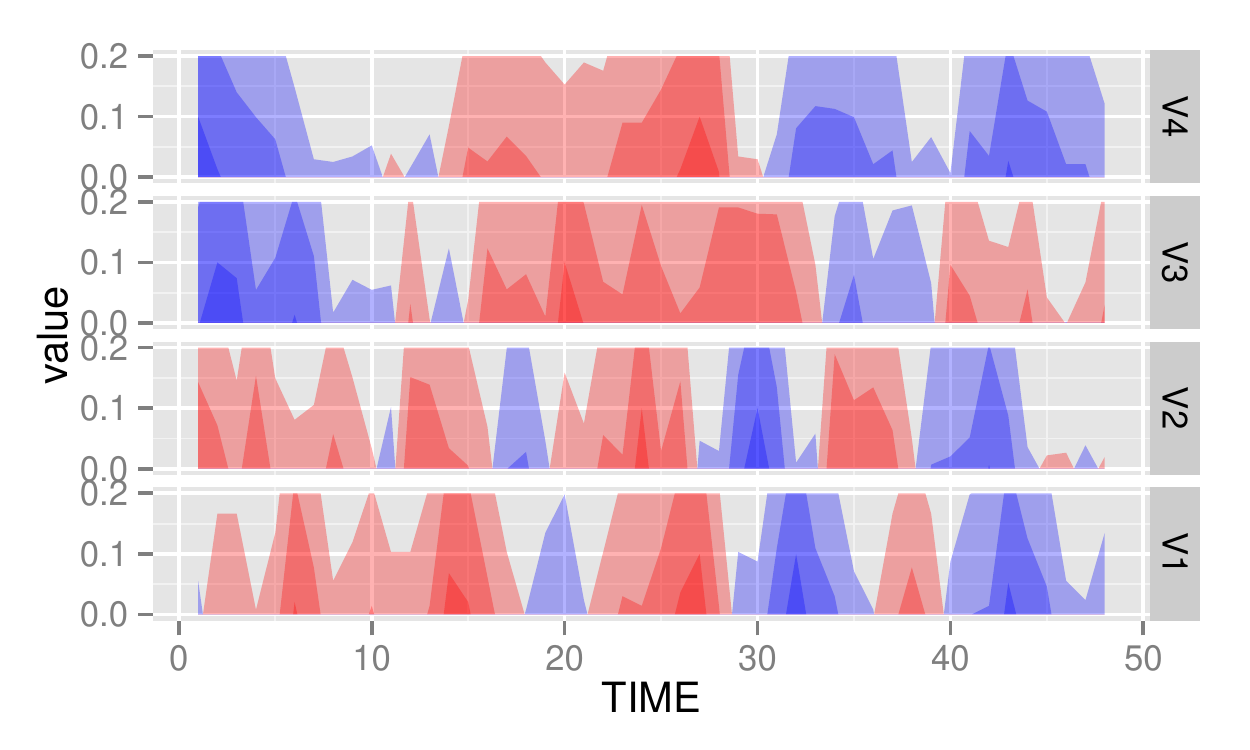}
\end{tabular}
\end{centering}
\caption{\label{fig:horizontal-axis} Six variations of horizontal axis time plots for multivariate
time series: (From top left to bottom right) overlaid line plots, faceted line plots,  stacked graph, faceted area chart, themeriver, horizon graph. Plots in the right column are examples of small multiples. }
\end{figure}
\end{center}

\end{itemize}

When time can be broken into two components it might be displayed
on both horizontal and vertical axes:

\begin{itemize} \itemsep 0in
\item A high frequency time series often has hierarchic or nested
period levels, like year, day, minute, etc. Those levels can
be placed on horizontal and vertical axes to reveal the periodic
dependency. For example, \citet{keller1993visual} used days
and hours on two axes. In these graphs, the measurements of
the time series are drawn in the grids via aesthetic settings
like color or size.

\item Calendar heat maps. \citet{van1999cluster} proposed a
colored calendar visualization  (weeks and days on two axes).
\texttt{\textbf{d3.js}} \citep{bostock2011d3} applies the calendar
heat maps and makes it interactive.
\end{itemize}

\begin{center}
\begin{figure}[htp]
\begin{centering}
\includegraphics[width=0.32\textwidth]{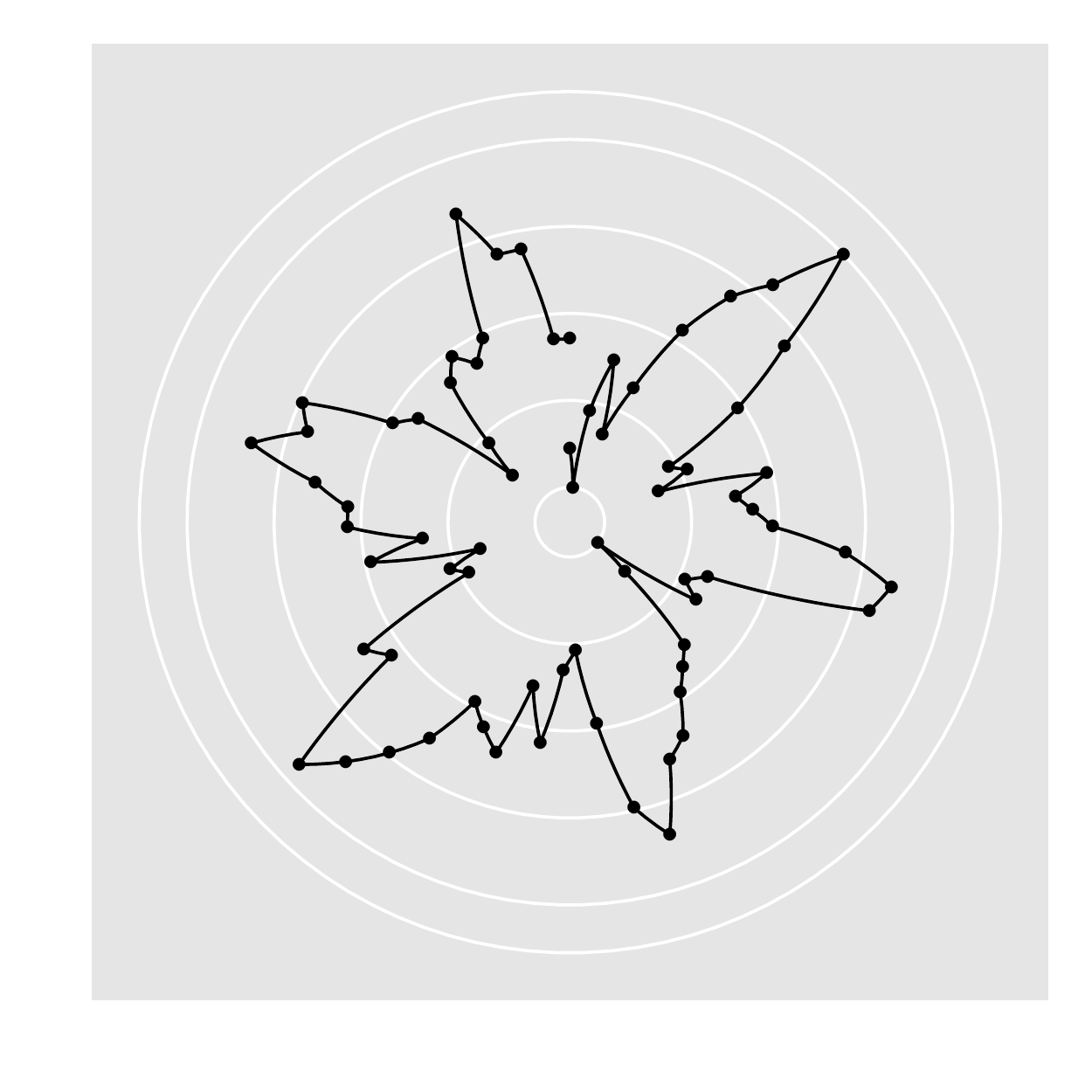}
\includegraphics[width=0.32\textwidth]{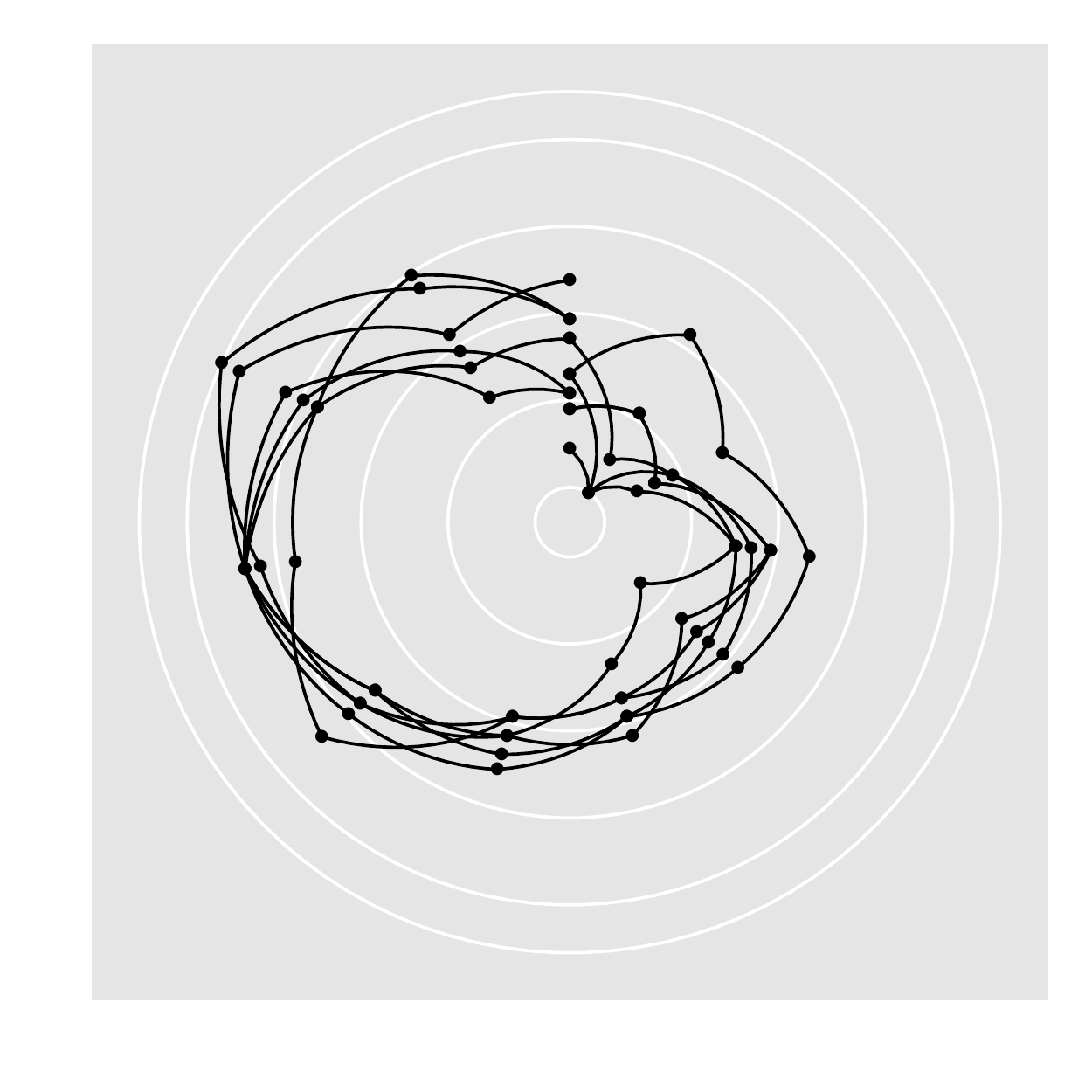}
\includegraphics[width=0.32\textwidth]{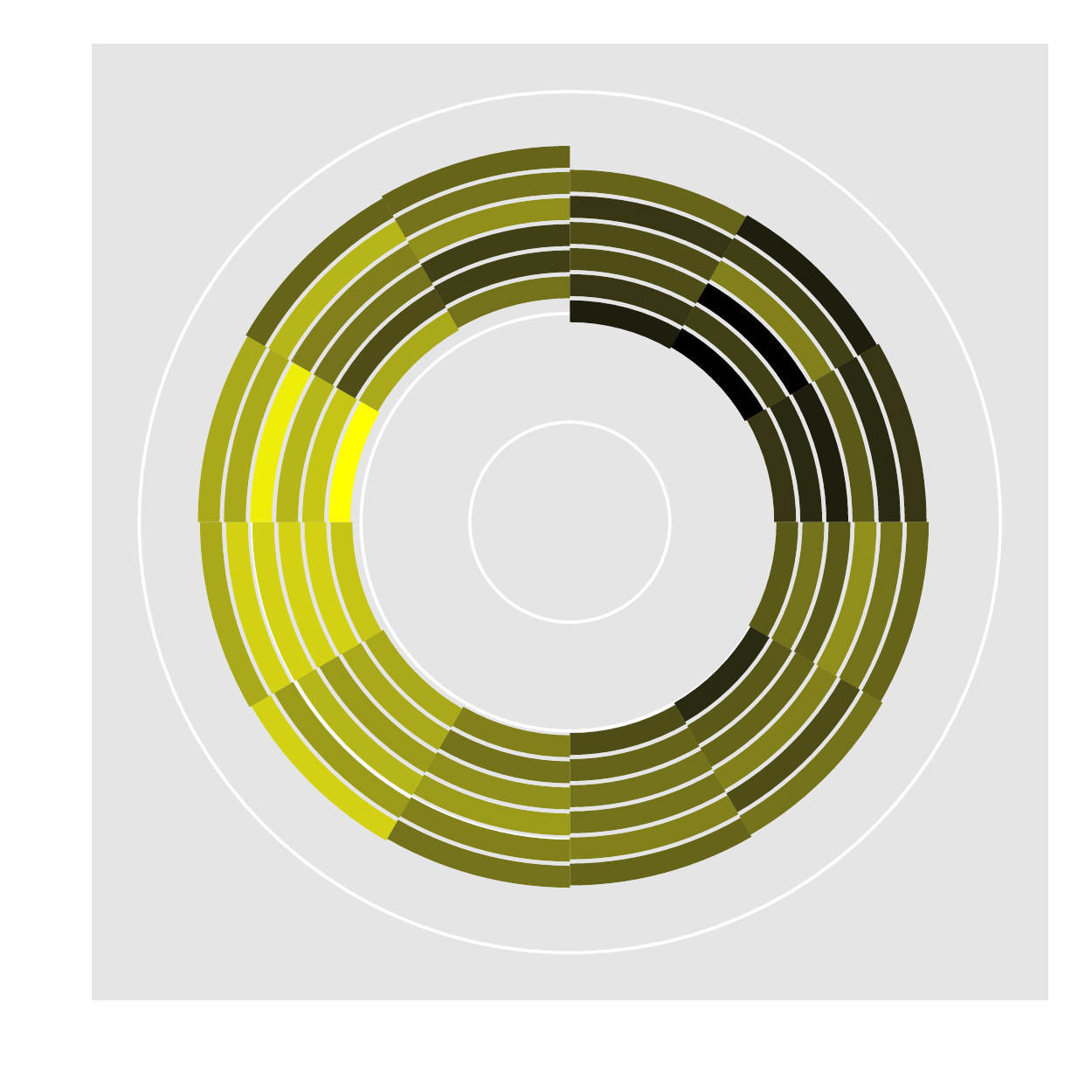}
\end{centering}
\caption{\label{fig:polar-axis}Three types of time series plots
in polar coordinates for the same data as used regularly spaced series shown in Figure
\ref{fig:Time-series-plots}:  direct conversion (left), wrapped by the period (middle), and
spiral graph with the colored grey scale representing the values (right).}
\end{figure}
\end{center}

Because time in some circumstances can be considered to be cyclical it is sometimes displayed in the polar coordinates:
 
\begin{itemize} \itemsep 0in
\item Nightingale's coxcomb. Florence Nightingale might be the earliest
author of a time series plot in polar coordinates. In the original
plots, two unstacked barcharts were made in polar coordinates. Each
diagram represents for one year. Later, people use the Nightingale's
coxcomb \citep{nightingale1858notes}, also called circular histogram
or rose diagram \citep{nemec1988shape}, to plot the time series with
a regular period like year or day.
\item Spiral graphs. This approach is proposed by \citet{weber2001visualizing}.
It can be seen as a temporal heatmap in polar coordinates. Figure 
\ref{fig:polar-axis} (right) shows an example. This approach
is good for seeking the period, but the length for the same time unit
changes over the loops. Besides, spiral graphs would be unhandy for
the short period problems and multiple time series.
\end{itemize}

\subsection{Interactive graphics\label{sec:interactive-graphics}}

Interactive graphics emphasize the user manipulation of plot elements via input devices like the keyboard and mouse \citep{symanzik2012interactive}. \citet{swayne1999} surveyed the use of the term ``interactive graphics'', which revealed some differences in what people mean when they use the term. They found that most commonly people perceived interactive graphics to mean that a new plot can be recreated quickly from the command line, like base R plots. They suggested using a different term, direct manipulation, to mean directly changing elements of the plot using input devices. However, this term did not gain traction in the community, and we still use interactive graphics. To be clear, we use it here to indicate direct manipulation of plot elements through input actions of mouse or key strokes. 

The work described here builds from a history of statistics software systems that support interactive graphics: e.g. 
\texttt{\textbf{PRIM-9}}~\citep{fisherkeller1988prim},
\texttt{\textbf{Data Desk}}~\citep{velleman1988datadesk},
\texttt{\textbf{LISP-STAT}}~\citep{tierney1990lisp},
\texttt{\textbf{XGobi}}~\citep{swayne1998xgobi} and
\texttt{\textbf{GGobi}}~\citep{cook2007ggobi},
\texttt{\textbf{MANET}}~\citep{unwin1996manet},
\texttt{\textbf{Mondrian}}~\citep{theus2002mondrian}.
The software \texttt{\textbf{Diamond Fast}}~\citep{unwin1988eyeballing},
\texttt{\textbf{XQz}}~\citep{McDougall1994}, and \texttt{\textbf{Fortune}}~\citep{kotterfortune}  provided tools specifically for exploring time series data.
With the current popularity of R language \citep{Rlanguage}, ideally interactive graphics can
integrate closely and flexibly with statistical modeling. Packages that support this to varying extents are \texttt{\textbf{rggobi}}~\citep{rggobi},
\texttt{\textbf{iplots}}~\citep{iplots},
\texttt{\textbf{rgl}}~\citep{adler2003rgl},
\texttt{\textbf{cranvas}}~\citep{cranvas},
\texttt{\textbf{ggvis}}~\citep{ggvis},
and \texttt{\textbf{animint}}~\citep{animint}. 

Of these software, \texttt{\textbf{cranvas}}, which evolved substantially from  \texttt{\textbf{GGobi}}, is the vehicle for the ideas described in this paper.
At its foundation is a data pipeline that channels data to
plot elements, and provides interaction through reactive data
elements, using \texttt{\textbf{plumbr}}~\citep{plumbr2014}. The
graphics are constructed using \texttt{\textbf{Qt}}~\citep{Qt}
that enables flexible plot design and fast rendering for smooth
interaction. \texttt{\textbf{cranvas}} has
many different types of plots and possible interactions.
The design of \texttt{\textbf{cranvas}}, \citet{xie2014reactive} provides single display interactions and the linking between different displays. Single display interactions include brushing, zooming, panning, and querying. Linked brushing can be done between different displays. To integrate temporal displays in this system requires integrating with this setup.

The next section describes the building blocks for temporal displays, which is followed by a taxonomy of interactive tasks that is desirable to use for exploration (Section \ref{sec:Design-of-interactions}). How to realize the interactions is described in Section \ref{sec:Pipeline}. Linking between temporal data displays and other plots is described in Section \ref{sec:Linking}.

\section{Layering to create a plot\label{sec:graph-layers}}

 For interactive graphics, layering up the plot to enable different interactions can be useful, and efficient for large data. The base layer is typically the plot of all of the data. An overlay of a brush layer, where only elements actively being colored are displayed, can provide the efficiency of faster rendering. A brush layer is common to all displays because brushing is a basic function for interactive graphics. Background layers like an axis layer or grid layer are common too, but are not necessary in displays like maps.  Table \ref{tab:Layers} lists the layers of common plot types in \texttt{\textbf{cranvas}}.

\begin{center}
\begin{table}[h]
\begin{center}
\begin{tabular}{c|l|l}
\hline 
  & scatterplot & point\tabularnewline
\cline{2-3} 
display specific & histogram & bar, cue\tabularnewline
\cline{2-3} 
 & map & polygon, googlemaps, path, point\tabularnewline
\cline{2-3} 
 & time plot & point, line, area, stats\tabularnewline
\hline 
common & required & brush, identify, keys\tabularnewline
\cline{2-3} 
 & optional & grid, $x$-axis, $y$-axis, $x$-label, $y$-label, title \tabularnewline
\hline 
\end{tabular}
\end{center}
\caption{\label{tab:Layers}Examples from \texttt{\textbf{cranvas}}
of layers used in constructing plots. Some are common to all plots,
and each plot has some layers that are unique. The ``keys'' layer
listens for key strokes that change interaction modes. The ``cue''
layer on the histogram contains listeners and handles for dragging
to change the binwidth interactively.}
\end{table}
\end{center}

For temporal data displays, there are three basic layers: point, line, and area. The coordinates of the points are initially calculated from the data, and interactions may change the locations of the points in the display. The line layer connects the current positions of the points, so it requires the order, or, path information to know how to make the connections. The area layer shades the area under the line by constructing a baseline, matching the minimum data values, which enables closing the series to create a set of polygons.
Each of these layers can take different interactions, and some care needs to be taken in realizing the effect on the different layers. 

The point attributes, selected, color, size, or visibility, are generated for each observation when creating the plumbr mutaframe. The base element for the temporal plots are the points, and in \texttt{\textbf{cranvas}} brushing changes the attribute of each point in the mutaframe -- essentially, points are brushed. The number of lines is one less than the number of points, and line color follows the first point in the defining pair. The number of polygons for the area display is the same as the number of lines, so color follows lines directly to polygons. The additional construction points of the area layer are only used in the area layer, and do not have independent attributes. This affects brushing behavior which is discussed later. 

\section{A taxonomy of interactions for temporal data displays\label{sec:Design-of-interactions}}

\citet{wills2012visualizing} summarizes
the interactivity for temporal displays as changes to parameters or data. Data is mapped into coordinates in the plot. Parameters can be considered to be attributes like color, labels, geometric elements, facet, or they can be considered to be aspects used to get the data into the plot like transformations, binning, dimension reductions or scales. Changes to the data or parameters provoke changes to the plots. 

Parameters can often be attached to graphical user interface (GUI) items like sliders, that can generate the change in the plot. But more generally, interactions happen by direct action on the plot. For example, in brushing, the user selects elements like points in the plots. The software needs to locate these items in the data, update the attributes of these selected points, and broadcast these changes to other plots. 

Some interactions, like brushing, selection, linking, zooming, panning, and querying, are universal for all plot types. Temporal and longitudinal data solicit special interactions to explore aspects of temporal dependence and trend.
\citet{Buja1996} describe a taxonomy of interactive tasks for multivariate data. Here we describe a taxonomy of tasks for temporal data, that enable exploration of different components of time series and longitudinal data:

\begin{figure}[htp]
\begin{center}
\begin{tabular}{cc}
\multicolumn{2}{c}{\includegraphics[width=0.8\textwidth]{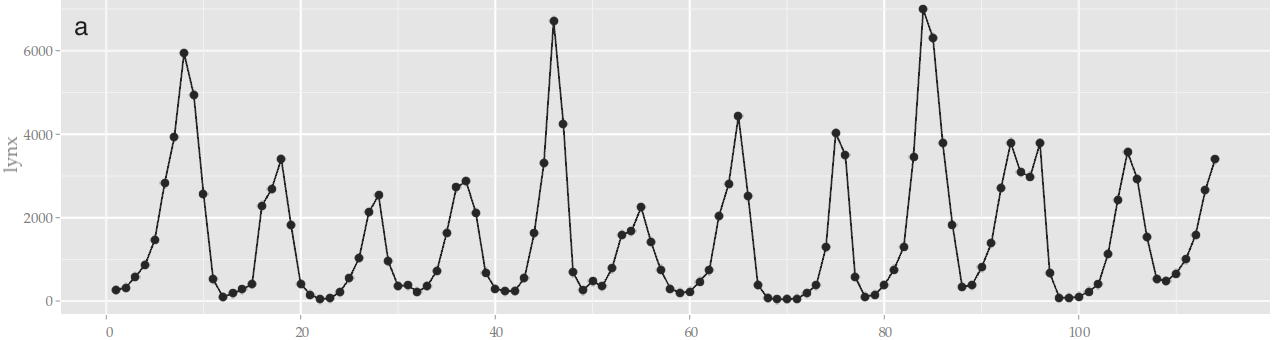}}\\
\includegraphics[width=0.4\textwidth]{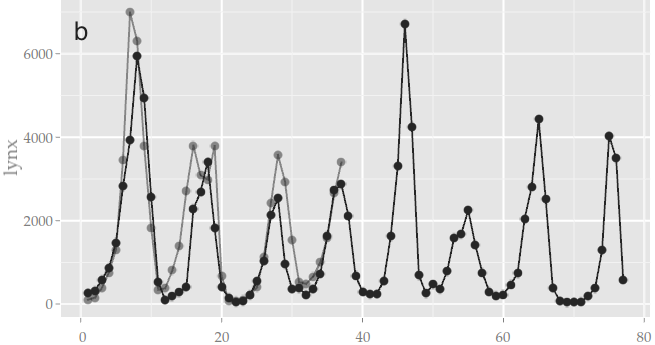} &
\includegraphics[width=0.4\textwidth]{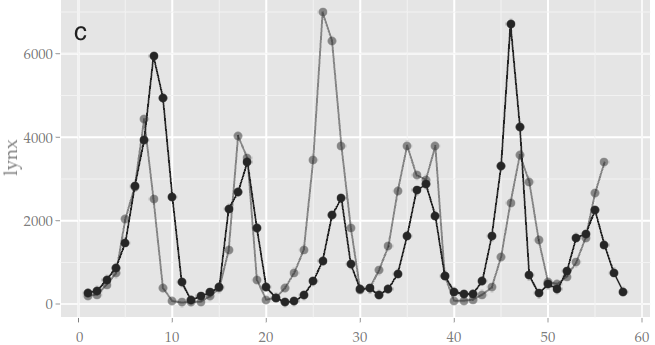} \\
\includegraphics[width=0.4\textwidth]{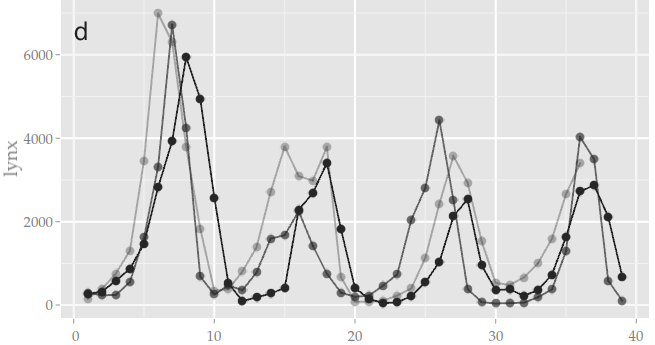} &
\includegraphics[width=0.4\textwidth]{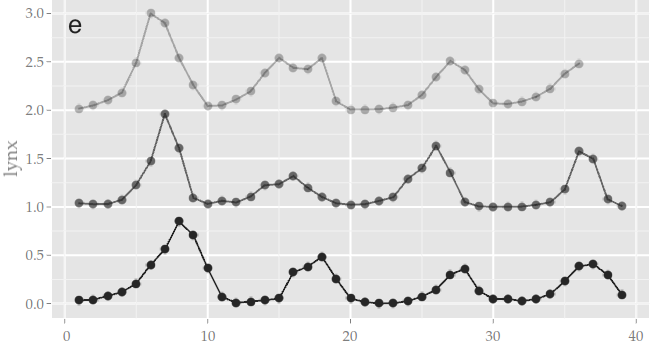} \\
\multicolumn{2}{c}{\includegraphics[width=0.8\textwidth]{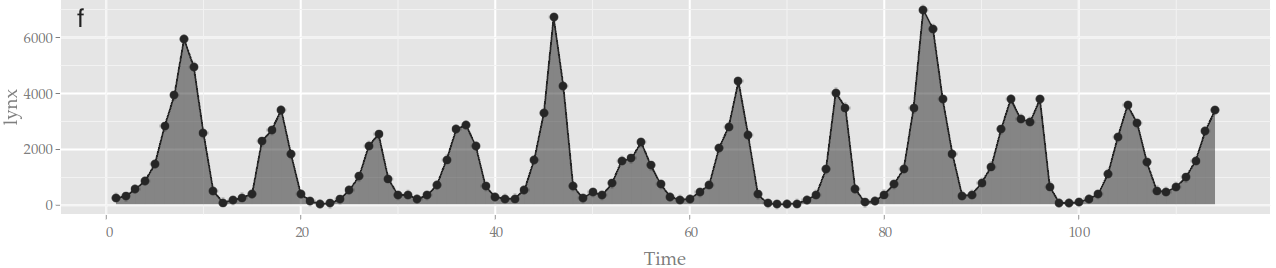}} \\
\multicolumn{2}{c}{\includegraphics[width=0.8\textwidth]{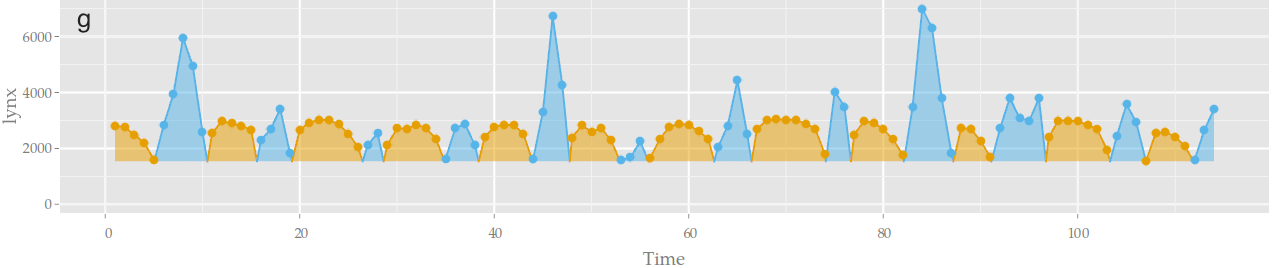}} \\
\end{tabular}
\caption{\label{fig:x-wrapping}Lynx trappings for 1821\textendash{}1934:
(a) Time series, (b \textendash{} d) stages of $x$-wrapping, matching
peaks, (e) faceted on the wrapped series, (f) area plot, (g) mirrored
on the mean, blue indicating values above the mean, and yellow below
the mean.  (Video illustrating these interactions is available at
\url{https://vimeo.com/112431547} and \url{https://vimeo.com/112432400}.)}
\end{center}
\end{figure}

\begin{center}
\begin{figure}[htp]
\begin{centering}
\includegraphics[width=0.98\textwidth]{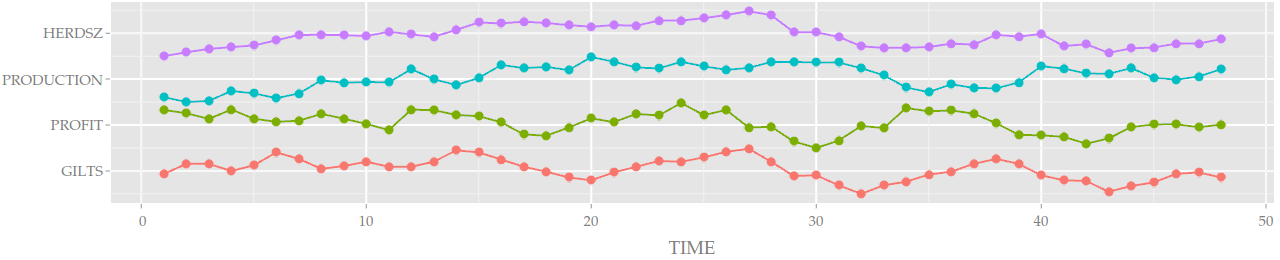}
\end{centering}
\begin{centering}
\includegraphics[width=0.98\textwidth]{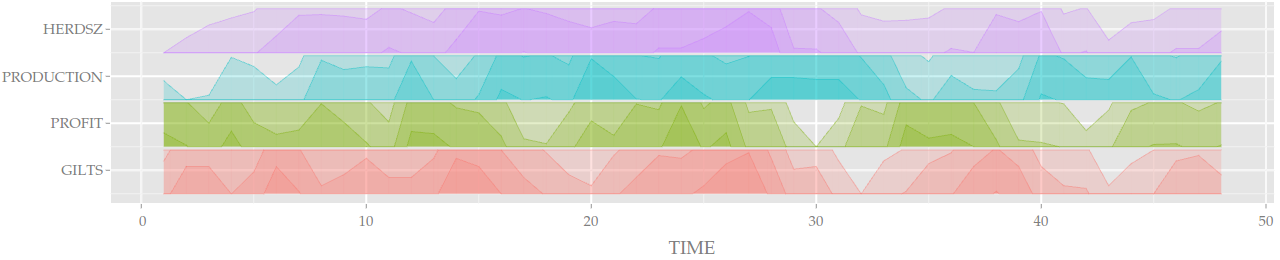}
\end{centering}
\caption{\label{fig:y-wrapping}Quarterly pig production for
1967-1978 in UK measured by five variables: (a) faceted, (b)
$y$-wrapped (see video at \url{https://vimeo.com/112435889}).
Profit and gilts seem slightly lag-related. Herdsz and
production may be related in a lag relationship also, but
neither is seasonal. In the $y$-wrapped version density indicates
the magnitude of values, and long periods of higher values,
like in herdsize are more visible. }
\end{figure}
\end{center}

\begin{center}
\begin{figure}[htp]
\begin{centering}
\begin{tabular}{cc}
\multicolumn{2}{c}{\includegraphics[width=0.48\textwidth]{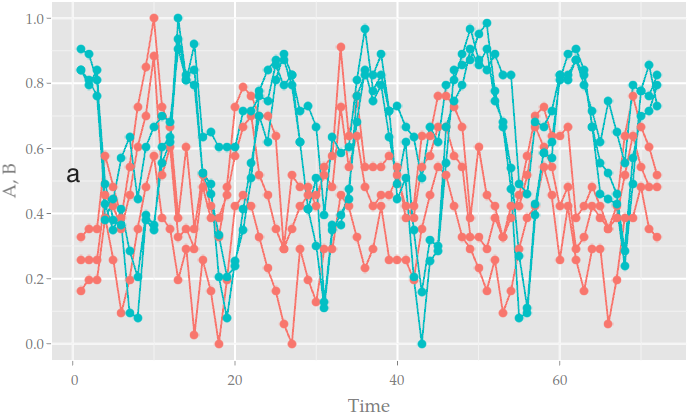}}\tabularnewline
\includegraphics[width=0.48\textwidth]{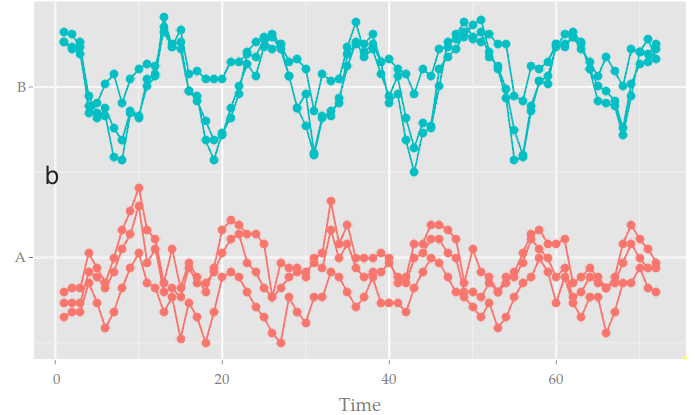} & \includegraphics[width=0.48\textwidth]{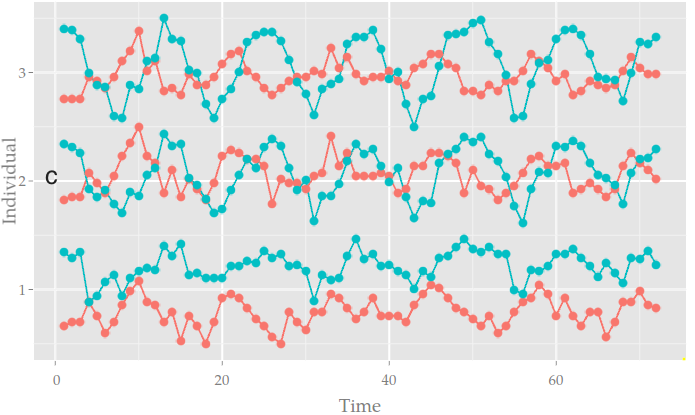}\tabularnewline
\includegraphics[width=0.48\textwidth]{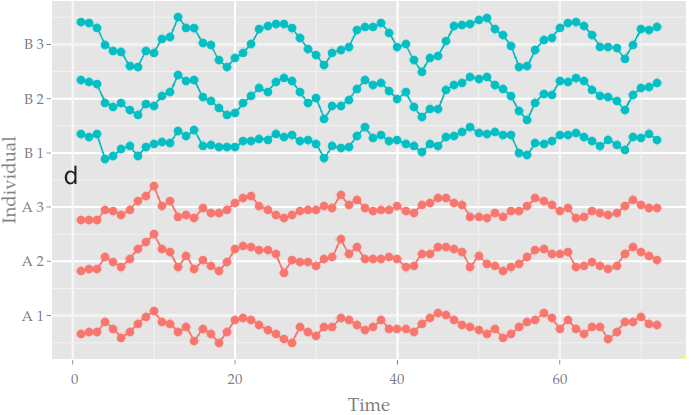} & \includegraphics[width=0.48\textwidth]{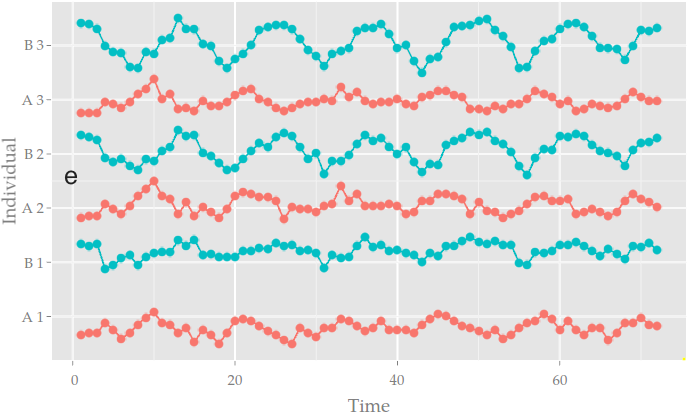}\tabularnewline
\end{tabular}
\end{centering}
\caption{\label{fig:faceting-var-ind}Order of interaction matters when faceting with two variables and three individuals: (a) all series overlaid, color indicates variable, (b) facet first on variable (A, B),  (c) facet first by individual (1, 2, 3), (d) facets (b) by individual, and (e) facets (c) by variable. Both final configurations are useful: (d) supports the primary comparison of individuals, with variable comparisons secondary and (e) supports comparison of variables within individuals.  (Video footage illustration is available at \url{https://vimeo.com/112438919}.)}
\end{figure}
\end{center}

\begin{figure}[htp]
\begin{center}
\includegraphics[width=0.45\textwidth]{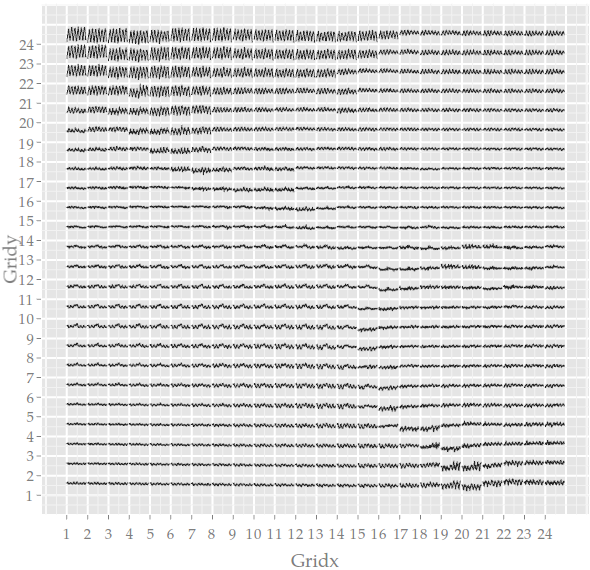}
\includegraphics[width=0.54\textwidth]{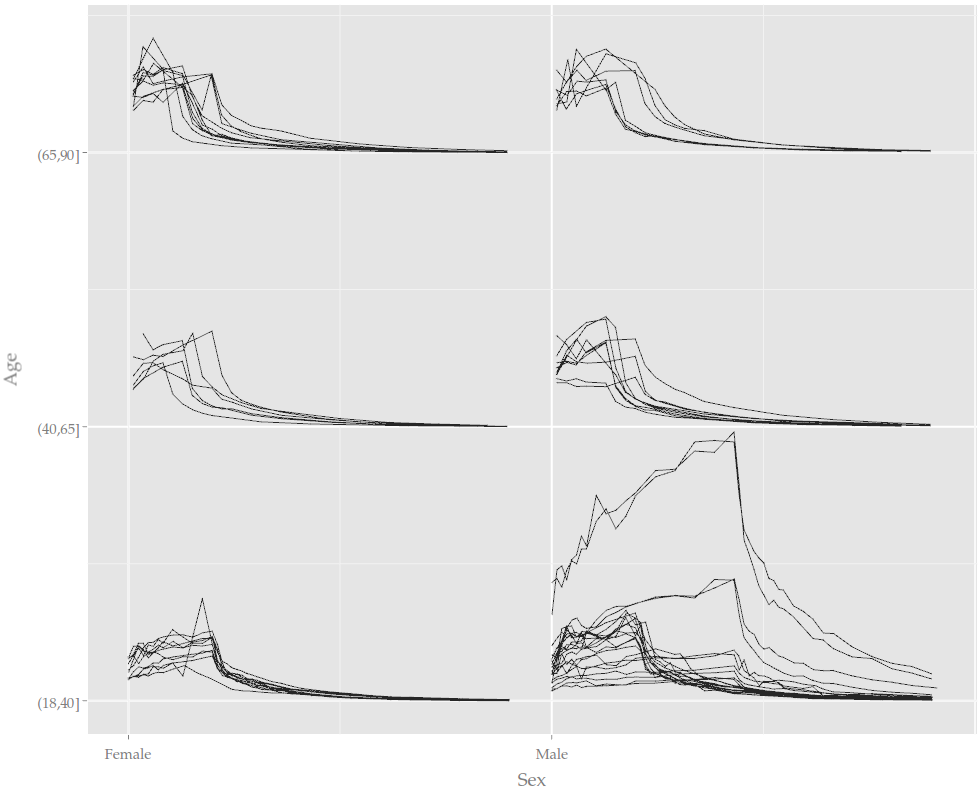}
\caption{\label{fig:faceting-examples}Faceting can be conducted bi-directionally: (left) by spatial grid for spatiotemporal data (\url{https://vimeo.com/112503285}), (right) by two covariates, sex and age, in longitudinal data (\url{https://vimeo.com/112509324}).}
\end{center}
\end{figure}

\begin{center}
\begin{figure}[htp]
\begin{centering}
\includegraphics[width=0.48\textwidth]{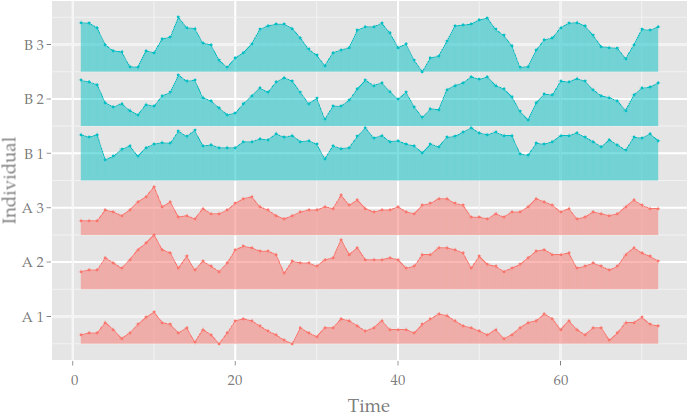}
\includegraphics[width=0.48\textwidth]{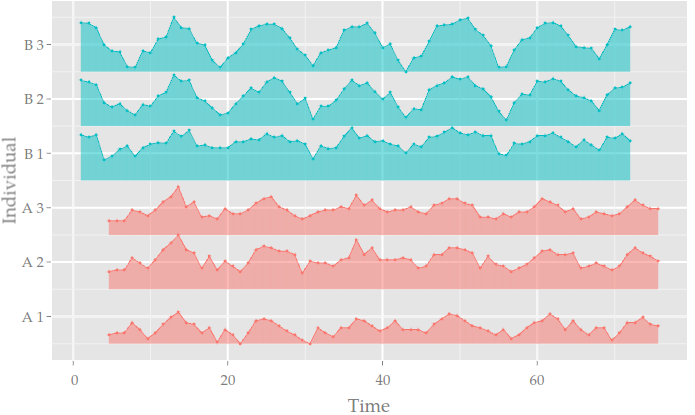}
\end{centering}
\caption{\label{fig:x-shifting}Illustration of shifting: three bottom series (variable A) are shifted horizontally to the right, to match the peak time with the top three series (variable B). (See also the video at \url{https://vimeo.com/112439923}.)}
\end{figure}
\end{center}

\begin{itemize} \itemsep 0in
\item Wrapping: In the $x$-direction explores seasonality and
  temporal dependence.  In the $y$-direction it is done to compare
  magnitude of peaks and dips. It is easiest to explain $x$-wrapping:
  the series is cut at a fixed-length interval, the part of the series
  that extends beyond this interval is re-drawn from the initial
  point. This will change the $x$-coordinates of the data. The main
  purpose of $x$-wrapping is to explore the regularity of the
  periodicity. Some series that look to follow a regular period can be
  quickly revealed to have irregularities. The classical example is
  the lynx trappings data for 1821\textendash{}1934 in the MacKenzie
  River District of North-West Canada\citep{campbell1977survey}, which
  looks periodic (Figure \ref{fig:x-wrapping}). The wrapping shows
  that the period is not quite regular, matching one peak off-sets
  other peaks, and the period varies between 9-11 years. It is also
  possible to see that the increases are slower than the drops, that
  the population builds up and tends to plummet. To model this data
  well requires one also knows the snowshoe hare population. Figure
  \ref{fig:y-wrapping} gives an example of the $y$-wrapping. It
  shows another classical example: quarterly pig production measured
  by four variables, herd size, production, profit and gilts, from
  1967-1978 in United Kingdom \citep{andrews1985data}. The $y$-wrapping
  induces something that might be considered a temporal boxplot, where
  density produced from overlaying the wrapped peaks emphasizes long
  runs of ups, or periodic ups.

\item Faceting: This creates small multiples to organize and examine across structural data components. These components might be a period such as year, or month, or variables when multiple are measured at the same time points, or in longitudinal data these might be individuals. In the interactive setting these components can be used to slowly pull overlaid series apart, a process that might be more revealing that disjointly laying out each in a static plot. Figure \ref{fig:faceting-var-ind} illustrates sequential faceting on two variables and three individuals. The order of these operations changes the final result, and changes which comparison is primary and which secondary. Figure \ref{fig:faceting-examples} gives two more examples, by a grid of spatial locations, and by covariates, respectively. Watch the videos to see how these are achieved interactively. There is also another video at \url{https://vimeo.com/112505175} shows faceting on period for a time series with a regular period of every 12 observations.

\item Mirroring: This splits the series vertically at a given value,
  and reflects the bottom half across this axis. With additional
  wrapping, the result is called a horizon graph, and it is used to
compare the magnitude of peaks and troughs, particularly for binary
phenomena like gains and losses. The $y$-coordinates of the data are
modified by this interaction. The choice of split value are typically
mean, median, midpoint of the range, or in economic data the initial
series value. Figure \ref{fig:x-wrapping} (f) and (g) shows mirroring
of the lynx trappings data with the mean divider. We can see the peaks
are sharp and irregular and the valleys are smooth and regular.

\item Shifting: A series can be grabbed and shifted against another series. This is a more tangible operation than wrapping in order to compare periodicity and temporal dependence. Figure \ref{fig:x-shifting} shows three series that have been picked up and shifted together against the other three series to match peaks. 

\item Switching: At any time it should be possible to switch between line and area displays. Line plots are efficient but filling the area under the curve can give a stronger sense of the patterns in the series, especially when trying to compare multiple series.   The video at \url{https://vimeo.com/112530645} demonstrates switching. 

\end{itemize}

\section{Display Pipeline\label{sec:Pipeline}}

\citet{wilkinson2000language, wilkinson2001nvizn, wilkinson2006grammar}
conceptualized and implemented a grammar of graphics that
carefully details a mapping of data to plots. It was extended
and implemented in ggplot2 by \citet{ggplot2}. Data is
parametrized into elements, and assigned to graphical elements,
e.g. points, text, lines, polygons. \citet{wills2012visualizing}
made a simple extension for interactive graphics: \textit{``allow
the user to manipulate one of the two inputs, data or parameters,
and show the changes in the chart''}. Parameters, very generally,
describe a very broad class of characteristics, e.g. display
aesthetics like color or linetype, positional coordinates,
statistics such as bins, scales like limits or color ladders,
facets, and transformations. Much of what is needed to realize
the taxonomy of tasks for interacting with time series plots
can be considered to be data transformations. This section
describes the transformations required to perform shifting,
faceting, wrapping, and mirroring.

Let $(\mathbf{x},\mathbf{y})$ denote the positional coordinates
for a temporal data set, where both are $n$-dimensional vectors
This notation is unconventional for time series, which typically
uses $x_t$ to represent the value records at time $t$, but it is
necessary for the graphical display because it allows us to think
about horizontal and vertical positions and adjustments to these
position. Because, many sequential interactions can be made, and
different types of interactions applied after each other, it is
useful to incorporate notation specifying these into the equations.
Let $\mathcal{I}$ be the temporal sequence of interactions,
e.g. $\{\textrm{facet},\textrm{wrap},\textrm{facet},
\textrm{zoom},\cdots\}$, and $j\in\mathcal{J}=\{1,2,\cdots,
J_{i}\}$ indicate the number of interactions made of type $i\in
\mathcal{I}$. Let $\mathbf{u}{}_{ij}=(u_{ij1},\, u_{ij2},\cdots)$
denote the user's input, e.g. key strokes, 
$\mathbf{l}{}_{ij}=(l_{ij1},\, l_{ij2},\cdots, l_{ijn})$
be a line group indicator for each point, since some interactions might force new sets of lines, 
$\mathbf{p}{}_{i}=(p_{i1},\, p_{i2},\, p_{i3},\cdots)$
be a parameter vector, e.g. a wrapping stop value of 3 points in series, and 
$\mathbf{m}{}_{ij}=(\Delta\mathbf{x}{}_{ij},\,\Delta\mathbf{y}_{ij})$
denote the movements in $x$- and $y$- directions, where $i\in\mathcal{I}$
is an interaction type, and $j$ is the number conducted. 
The new data coordinates are given by 
\[
(\mathbf{x},\mathbf{y})_{s+\mathcal{I}_{ij}}=(\mathbf{x},\mathbf{y})_{s}+\mathbf{m}{}_{ij},
\]
where $s$ indicates the state before interaction.
The movement $\mathbf{m}{}_{ij}$ can be written as a function on $\mathbf{p}{}_{i}$,
$\mathbf{u}{}_{ij}$, $\mathbf{l}_{ij}$, $j$, and the inital
coordinates $(\mathbf{x},\mathbf{y})_0$, 

\[
\mathbf{m}{}_{ij}=f_{i}(\mathbf{p}{}_{i},\:\mathbf{u}{}_{ij},\;\mathbf{l}_{ij},\; j,\; (\mathbf{x},\mathbf{y})_{0} ).
\]

\noindent These are the specific functional definitions used to generate the movement for the interactions available in \texttt{\textbf{cranvas}}. 

\subsection{Wrapping}

Figure \ref{fig:x-wrapping} (except bottom right plot)
illustrates the horizontal wrapping of the lynx trapping data.
The default wrapping interaction, by clicking a keystroke,
induces the point at the end of the series, $x_{(n)}$, to be
cropped and moved to the very left side of the plot, at the
same $x$-position as $x_{(1)}$. With repeated keystrokes, the
most recent elements of the series will be cropped and gradually
wrapped onto the earliest elements. Only the $x$-coordinates
are changed -- the $y$-coordinates remain unchanged.

For simplicity, we assume that the difference between consecutive
time values is 1. Let  $x_{(1)},\cdots,x_{(n)}$ be the sorted
$x$-coordinates of the $n$ points in the series, that is the
points in time order. The $x$-limits after $j$ keystrokes for
$x$-wrapping will be reset to $(x_{(1)}, x_{(n-j)})$, so the
plot is rescaled accordingly. Let $\Delta_{n-j}=x_{(n-j)}-x_{(1)}+1$,
then the new $x$-coordinate, $x^*$ of $x$ is
\begin{eqnarray*}
x^* & = & \begin{cases}
x_{(n-j)}  & \mbox{if~} x-x_{(1)}+1 \mbox{~mod}\Delta_{n-j} = 0\\
(x-x_{(1)}+1) \mbox{~mod}\Delta_{n-j} ~~+x_{(1)}-1 &  \mbox{o.w.}
\end{cases} \\ 
& = & \begin{cases}
x_{(n-j)}  & \mbox{if~} (x-x_{(1)}+1) \mbox{~mod}\Delta_{n-j} = 0\\
(x-x_{(1)}+1)-\left\lfloor\frac{x-x_{(1)}+1}{\Delta_{n-j}}\right\rfloor\times\Delta_{n-j}+x_{(1)}-1 &\mbox{o.w.}\\
\end{cases} \\
 & = &
(x-x_{(1)}+1)-\left(\left\lceil\frac{x-x_{(1)}+1}{\Delta_{n-j}}\right\rceil -1\right)\times \Delta_{n-j} +x_{(1)}-1 \\ & = &
x-\left(\left\lceil \frac{x-x_{(1)}+1}{\Delta_{n-j}}\right\rceil -1\right)\times\Delta_{n-j},
\end{eqnarray*}
enabling the movements for $i=$ wrap to be described as
\begin{eqnarray*}
\mathbf{m}{}_{ij} & = & \begin{cases}
(-\left(\left\lceil \frac{\mathbf{x}-x_{(1)}+1}{\Delta_{n-j}}\right\rceil -1\right)\times\Delta_{n-j}, \; 0) & 1\leq j \leq n-3 \\
(-\left(\left\lceil \frac{\mathbf{x}-x_{(1)}+1}{\Delta_3}\right\rceil -1\right)\times\Delta_3, \; 0) & j\ge n-2
\end{cases} \\ 
\end{eqnarray*}
or equivalently in terms of line group indicators as well as points as
\begin{eqnarray*}
\mathbf{m}{}_{ij} & = & \begin{cases}
(-(\mathbf{l}{}_{ij} -1)\times\Delta_{n-j}, \; 0) & 1\leq j \leq n-3 \\
(-(\mathbf{l}{}_{ij} -1)\times\Delta_3, \; 0) & j\ge n-2.
\end{cases}
\end{eqnarray*}

The line group indicator $\mathbf{l}{}_{ij}$ will depend on the number of interactions $j$. The wrapping can be defined as an algorithm also:

\begin{enumerate} \itemsep 0in
\item Shift the data values up, usually by 1
\item Check the new $x$-limits, if a point has value large than upper limit, crop it using modulus arithmetic
\item Points that are cropped, have their line group indicator incremented
\item Connect the points that have the same line group indicator, in time order.
\end{enumerate}

Sometimes it is useful to wrap the series faster. If you have
a long time series, it might be useful to make full year jumps.
This can be achieved with the above equations by setting the
sequence of $j$ to respect this period. In other instances it
may be useful to have a multiplicative wrapping so that it
looks like the series wraps faster and faster with each step.
That means every keystroke will send a different number of
points from the right to the left. The number of points
wrapped by the $j$th step, can be represented by the user
input parameter $u_{ij}$. Then the $x$-range after $j$ steps
is $(x_{(1)}, x_{(n-\sum_{a=1}^j u_{ia})})$, yielding
$\Delta_{n-\sum_{a=1}^j u_{ia}}=x_{(n-\sum_{a=1}^j u_{ia})}-x_{(1)}+1$, and
\begin{eqnarray*}
x^* & = & x-\left(\left\lceil \frac{x-x_{(1)}+1}{\Delta_{n-\sum_{a=1}^j u_{ia}}}\right\rceil -1\right)\times\Delta_{n-\sum_{a=1}^j u_{ia}}, \\
\mathbf{m}{}_{ij} & = & \begin{cases}
(-(\mathbf{l}{}_{ij} -1)\times\Delta_{n-\sum_{a=1}^j u_{ia}}, \; 0) & 1\leq \sum_{a=1}^j u_{ia} \leq n-3 \\
(-(\mathbf{l}{}_{ij} -1)\times\Delta_3, \; 0) & \sum_{a=1}^j u_{ia}\ge n-2.
\end{cases}
\end{eqnarray*}

If the user wants to skip all intermediate positions and use
only one jump to the fully wrapped position,
then the new $x$-range will be $(x_{(1)}, x_{(p_{i2})})$, where
the parameter $p_{i2}$ is the length of period. Hence for $j\ge 1$,
\begin{eqnarray*}
x^* & = & x-\left(\left\lceil \frac{x-x_{(1)}+1}{\Delta_{p_{i2}}}\right\rceil -1\right)\times\Delta_{p_{i2}}, \\
\mathbf{m}{}_{ij} & = &
(-(\mathbf{l}{}_{ij} -1)\times\Delta_{p_{i2}}, \; 0).
\end{eqnarray*}

To generalize our case to the irregular time series, we
should specify a wrapping speed parameter $p_{i3}$, i.e.,
with every key stroke, the $x$-range is shortened by at
least $p_{i3}$. The wrapping speed parameter will determine
how many points are shifted every time, because if the
difference between largest two points is greater than
$p_{i3}$, then only one point is shifted; if the difference
is smaller than $p_{i3}$, then more than one points are
shifted. After $j$ steps, the total number of points
shifted is a function of $p_{i3}$ and $\mathbf{x}_0$.
Denote the function as $g_j$, so the new $x$-range is
$(x_{(1)},x_{(n-g_j(p_{i3},\mathbf{x}_0)})$, and the
movements can be calculated then.

Movements from the $y$-wrapping on the $y$-direction,
as shown in Figure \ref{fig:y-wrapping}, could be obtained
by similar formulas. It is messier to realize because the
$y$-values are typically not in a sequential order which
means that more structural components need to be added to
the data to actually draw the wrapped series. Some of the
issues are discussed later in this paper.

\begin{center}
\begin{figure}[htp]
\centerline{\includegraphics[width=0.9\textwidth]{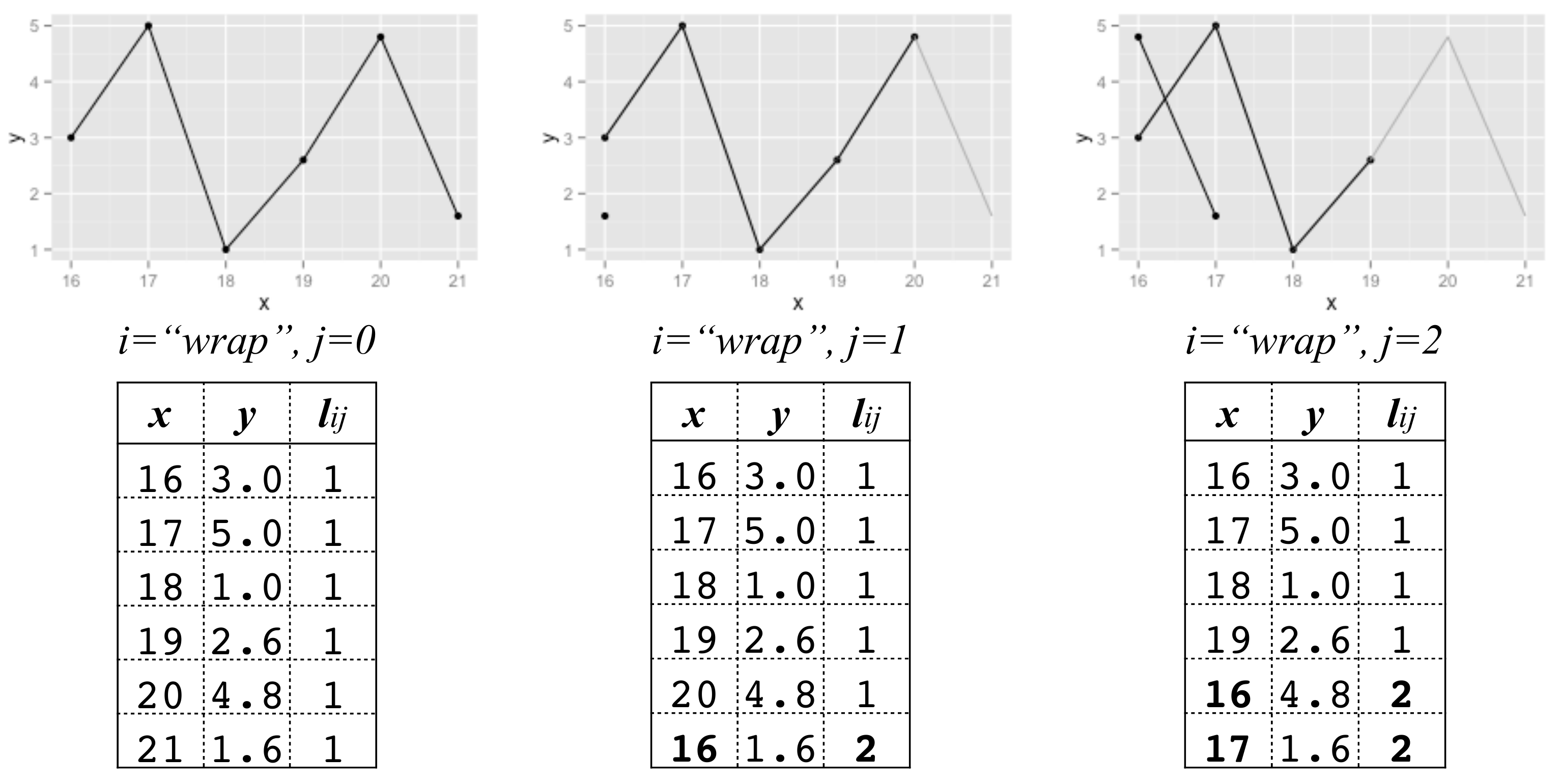}}
\caption{\label{fig:x-wrapping-algorithm} Cartoon illustrating
the $x$-wrapping. There are three consecutive wrapping
steps. At step $j=1$ the $x$ value of the last point changes
from 21 to 16, and the line group indicator increments to 2.
At step $j=2$ the $x$ values of the last two points are changed,
and both have line group indicators equal to 2.  The wrapping
stop parameter was set to $p_{i1}=3$, which means that after
one more step, $n-j=3$, the wrapping would stop because each
series has only 3 points. In the actual implementation the
scale of the horizontal axis is changed at each step so that
the full plot width is used.}
\end{figure}
\end{center}

\subsection{Faceting}

When $i=$ facet by individual, an initial setting of the
parameter $p_{i1}=0.05$, which means that every hit on the
key will lift the $l$th standardized line by $(l-1)\times0.05$.
Hence for $j\in\mathcal{J}$,
\begin{eqnarray*}
\mathbf{m}{}_{ij} & = & \begin{cases}
 (0,\;0.05\, (\mathbf{l}{}_i-1)\, j) & 1\leq j<20,\\
 (0,\; \mathbf{l}{}_i-1) & j\ge20.
\end{cases}
\end{eqnarray*}
We can also generalize the equation above by
\begin{eqnarray*}
\mathbf{m}{}_{ij} & = & \begin{cases}
 (0,\; p_{i1}\, (\mathbf{l}{}_i-1)\, j) & 1\leq j<\frac{1}{p_{i1}},\\
 (0,\; \mathbf{l}{}_i-1) & j\ge\frac{1}{p_{i1}},
\end{cases}
\end{eqnarray*}
where $p_{i1}\in (0,1)$.

The example shows that $\mathbf{m}{}_{ij}$ is a function of $\mathbf{p}{}_{i}$,
$j$, and $\mathbf{l}_{ij}$, where $\mathbf{l}_{ij}=\mathbf{l}{}_i$ in this example
means that the line indicator for faceting is free from $j$.

For $i=$ facet by variable/period, one click will fully split
the variables, so $j$ does not matter. All lines should be 
standardized between $[0,1]$ first, then the movement is given by
\[
\mathbf{m}{}_{ij} = (0,\; \mathbf{l}{}_i-1).
\]
Note that $\mathbf{l}{}_i$ in this case differs from
$\mathbf{l}{}_i$ in faceting by individual.

\subsection{Mirroring}

To realize the interaction shown in Figure \ref{fig:x-wrapping} (g),
firstly we need to point the divider -- mean in this example.
Hence for $i=$ mirroring, the divider parameter
$p = \frac{1}{n}\sum_{d=1}^{n}y_d$. Then by $j\in\mathcal{J}$
hits on some triggering key, the movements are
\begin{eqnarray*}
\mathbf{m}{}_{ij} & = & \begin{cases}
(0, \; p+\max(p-\mathbf{y},\,\mathbf{y}-p)-\mathbf{y}) & j=1,3,5,\cdots \\
(0,0) & j=2,4,6,\cdots
\end{cases}\\ & = & \begin{cases}
(0, \; \max(2p-2\mathbf{y},\,0)) & j=1,3,5,\cdots \\
(0,0) & j=2,4,6,\cdots
\end{cases}
\end{eqnarray*}
Note that if the mirroring is revisited after some other interactions,
then the count of $j$ should not be reset.

\subsection{Shifting} 

Figure \ref{fig:x-shifting} illustrates shifting the series, which is used to compare one series against another. The user input uses $\mathbf{u}{}_{ij}$, since the user can drag the
series horizontally to any position. The starting point
$u_{ij1}$ and end point $u_{ij2}$ of dragging on the $x$-axis,
as well as the selected series $u_{ij3}$ are the input from the user.
The horizontally shifting will not change $y$-coordinates, so for
$i=$ $x$-shifting and $j\in\mathcal{J}$, we have
\begin{eqnarray*}
\mathbf{m}{}_{ij} & = & 
((u_{ij2}-u_{ij1})\times I\left\{ \mathbf{l}{}_{ij}=u_{ij3}\right\}, \; 0),
\end{eqnarray*}
where $I$ is the indicator function.

\subsection{Additivity of interactions\label{interaction-addition}}

Most of the interactions could be considered to be additive.
Figure \ref{fig:faceting-var-ind} shows an example, where two
different results are generated by different ordering of interactions.
Faceting on individual is done after faceting on variable, with the
process following panels (a) $\rightarrow$ (b) $\rightarrow$ (d).
Faceting on variable after individual, as in the process (a)
$\rightarrow$ (c) $\rightarrow$ (e), produces a different
configuration of the time series. The additive application of
interactions is not commutative. Both results are useful, because
each facilitates a different type of comparison of the series,
using proximity. Plot (d) enables the comparison of individuals,
within variables, while plot (e) enables the comparison of series
within individual. It is also interesting to
note that wrapping vertically after mirroring, will result in
a horizon graph, like Figure \ref{fig:horizontal-axis} (f).

The cumulative interactions could
entirely change both $x$ and $y$ coordinates of data.
For example, Figure \ref{fig:x-wrapping} firstly directs
75 steps of $x$-wrapping, and then a faceting by
period. That gives the eventual movement by
\begin{eqnarray*}
\mathbf{m} & = & (-\left(\left\lceil \frac{\mathbf{x}-x_{(1)}+1}{\Delta_{39}}\right\rceil -1\right)\times\Delta_{39}, \; \mathbf{l}{}_{facet}-1) \\
& = & (-(\mathbf{l}{}_{wrap,75} -1)\times\Delta_{39}, \; \mathbf{l}{}_{facet}-1) \\
& = & (-(\mathbf{l}{}_{wrap,75} -1)\times\Delta_{39}, \; \mathbf{l}{}_{wrap,75}-1).
\end{eqnarray*}

Note that $\mathbf{l}{}_{facet} = \mathbf{l}{}_{wrap,75}$
in this example, because the line group indicator for
faceting by period is given by the wrapping steps.

In some other cases, a combination of interactions may
only modify either $x$ or $y$ coordinates. Figure
\ref{fig:faceting-var-ind} (a $\rightarrow$ b $\rightarrow$ d)
shows a combination of faceting first by variable then
by individual. The final movement after the full split
of individuals would be
\[
\mathbf{m} = (0, \; (\mathbf{l}{}_{facet~by~variable}-1)\times \max(\mathbf{l}{}_{facet~by~individual})+(\mathbf{l}{}_{facet~by~individual}-1)).
\]
Table \ref{tab:additive-faceting} takes the first point
of each series as an example to show $y$ and the changes
at the three stages.

\begin{table}[h]
\begin{center}
\begin{tabular}{|c|c|cc|c||c||c|}
\cline{1-2} \cline{5-7}
$l_{variable}$ & $l_{individual}$ &  &  & $y$ at (a) & $y$ at (b) & $y$ at (d)\tabularnewline
\cline{1-2} \cline{5-7}
1 & 1 &  &  & 0.16 & 0.16+(1-1)=0.16 & 0.16+(1-1)$\times$3+(1-1)=0.16\tabularnewline
\cline{1-2} \cline{5-7}
1 & 2 &  &  & 0.33 & 0.33+(1-1)=0.33 & 0.33+(1-1)$\times$3+(2-1)=1.33\tabularnewline
\cline{1-2} \cline{5-7}
1 & 3 &  &  & 0.26 & 0.26+(1-1)=0.26 & 0.26+(1-1)$\times$3+(3-1)=2.26\tabularnewline
\cline{1-2} \cline{5-7}
2 & 1 &  &  & 0.84 & 0.84+(2-1)=1.84 & 0.84+(2-1)$\times$3+(1-1)=3.84\tabularnewline
\cline{1-2} \cline{5-7}
2 & 2 &  &  & 0.84 & 0.84+(2-1)=1.84 & 0.84+(2-1)$\times$3+(2-1)=4.84\tabularnewline
\cline{1-2} \cline{5-7}
2 & 3 &  &  & 0.90 & 0.90+(2-1)=1.90 & 0.90+(2-1)$\times$3+(3-1)=5.90\tabularnewline
\cline{1-2} \cline{5-7}
\end{tabular}
\end{center}
\caption{\label{tab:additive-faceting}$y$-coordinates of the
first point on each line, at Figure \ref{fig:faceting-var-ind}
(a) no faceting, (b) faceting by variable, (d) faceting by
variable then individual.}
\end{table}

\subsection{Incremental vs baseline operations\label{sub:Two-procedures}}

Calculations can be made incrementally or with respect to a stable state (baseline), which, respectively, stores multiple copies of data, or a single storage of the data with storage of movement.
\begin{enumerate} \itemsep 0in
\item Incremental: Let

\begin{eqnarray*}
s_{0} & = & \textrm{initial status},\\
s_{t+1} & = & s_{t}+u_{i1}.
\end{eqnarray*}
 Note that every new status is only one interaction after the previous
status. The coordinates are given by
\begin{eqnarray*}
 (\mathbf{x},\mathbf{y})_{s_{t+1}} & = &  (\mathbf{x},\mathbf{y})_{s_{t}}+\mathbf{m}{}_{i1}\\
\mathbf{m}{}_{i1} & = & f_{i} (\mathbf{p}{}_{i},\;\mathbf{u}{}_{i1},\;\mathbf{l}_{i1},(\mathbf{x},\mathbf{y})_{s_0})
\end{eqnarray*}

This procedure always computes the next position directly from the
current position. The change only depends on the corresponding interaction
parameters and the input. The current status is stored in the memory
and ready to use for the next step. This method is an intuitive
design for interactive graphs with few special
interaction types. For example, scatterplots in \texttt{\textbf{cranvas}}
can change the size and transparency of dots. The new size (or
transparency) is always calculated by the multiplication of the
current size (or transparency) and a constant. The constant is
greater than 1 if the aesthetic parameter is increasing, and
less than 1 if the parameter is decreasing. The exponential
growth of the parameter accelerates the change and reduces the
times of repeated interaction.

The advantage of this procedure includes the straightforward design,
the convenience of moving to the previous or next status, and the
efficiency of avoiding the recomputation. However,
when there are many special interactions that could transform the
data, we need to record both the initial and current data positions
of each interaction type in the stream $\mathcal{I}$, because
when moving backwards, we need to know when the initial state is
reached and then stop. Then for an interaction stream $\mathcal{I}$
of length $k$, at least $k+1$ phases $ ( (\mathbf{x},\mathbf{y})_{s_{0}}, (\mathbf{x},\mathbf{y})_{s_{t_{1}}}, (\mathbf{x},\mathbf{y})_{s_{t_{2}}}\cdots, (\mathbf{x},\mathbf{y})_{s_{t_{k}}})$
should be saved, where $t_{1},t_{2},\cdots,t_{k}$ are the time of
the end of interaction types $1,2,\cdots,k$. When the data set is
large, those copies will occupy too much memory. Also, the storage
and management of $\mathbf{u}{}_{i1}$'s and $\mathbf{l}_{i1}$'s
is messy. Another drawback is that numerical errors could be
introduced after the same number of forward and backward
interactions, due to the floating-point arithmetic calculation.

\item Baseline: To store the $k+1$ phases, we do not make $k+1$
copies of the data set, instead, the movement item is traceable. The
coordinates of any status can be computed by 
\begin{eqnarray*}
 (\mathbf{x},\mathbf{y})_{s_{t}} & = &  (\mathbf{x},\mathbf{y})_{s_{0}}+\sum_{i,j}\mathbf{m}{}_{ij}\\
 & = &  (\mathbf{x},\mathbf{y})_{s_{0}}+\sum_{i,j}f_{i} (\mathbf{p}{}_{i},\;\mathbf{u}{}_{ij},\;\mathbf{l}_{ij},\; j,\; (\mathbf{x},\mathbf{y})_{s_0}).
\end{eqnarray*}
Note that when a new position is required, the calculation starts
from the initial position instead of the previous status. The movements
from the original to the current position are computed instantly.

The idea of baseline operation is not new. In the tour movement
of \texttt{\textbf{XGobi}} and \texttt{\textbf{GGobi}}, the target
position is calculated by the initial position and the projection
parameters that come from the auto-oriented settings or
user-oriented interactions \citep{cook1995grand, cook1997manual}.
However, the interactivities that we discussed in this paper is
more complex, because we need to consider the interaction stream
$\mathcal{I}$, but the tour movement does not need to.
When there is only one type of modification, the formulas in
Section \ref{sec:Pipeline}
will provide the target position easily.
 But when different types of modifications
are mixed, the baseline method structures the computation well.

With this procedure we do not need to save the intermediate
positions of the data, but we have to save the inputs for movements.
Now the problem turns to: how to store the inputs including
$\mathbf{p}{}_{i}$, $\mathbf{u}{}_{ij}$, $\mathbf{l}_{ij}$, and $j$? The answer is,
to save $\mathbf{l}_{iJ_i}$ with the data, where $J_i$ is the largest
$j$ in each $i \in \mathcal{I}$. This is because $\mathbf{p}{}_{i}$
is fixed and $j$ is known, $\mathbf{u}{}_{ij}$ is usually a short array,
but $\mathbf{l}_{ij}$ is of the same length as the data, and depends
on other parameters like $\mathbf{u}{}_{ij}$. So the data frame that
we use to save the data includes not only the coordinates, point parameters
like size and color, but also the line group indicators $\mathbf{l}_{iJ_i}$.

The advantage of this procedure is apparent: we do not have
to save multiple copies of data, and it is a better way to manage
the data and parameters during the interactions. However, it is not a comprehensive solution. We assumed that the movements are additive, but this is not always desirable. When changes in type of interaction make calculations better performed on a mid-way state, then it is better to stop, use this state as the baseline and then continue adding movements to this state.

\end{enumerate}

\section{Linking\label{sec:Linking}}

Linking between plots is a critical component of using multiple linked
windows \citep{stuetzle1987plot} to explore data. \citet{xie2014reactive} describes
types of linking and how it is realized in \texttt{\textbf{cranvas}}.
It is possible to both self-link, which is important for temporal data,
and link on different data sources or aggregation levels, using
categorical variables. For the temporal and longitudinal data,
linking is complicated when there is the need for different forms of
the dataset or additional data. Two situations are discussed in
the following sections.

\subsection{Self-linking}

Self-linking is primarily used to highlight all of the points in a time series when any one is selected. It is the most common behavior that a user would use. When there are multiple time series, it may also be useful to link to all points representing values recorded at a particular time. 

Data underlying multiple time series, as for most of the other
plots available in \texttt{\textbf{cranvas}}, are usually in
``wide data'' format (Table \ref{tab:wide-data}). One row contains
the values recorded for a particular time, and aesthetic
parameters are associated with each row. In this form if the
display shows the multiple series, then when a user selects
one point by brushing, all the points (values for V1, V2, V3)
for this time are highlighted. This form is not conducive to
selecting either a single point or an entire time series. 

\begin{figure}[htp]
\begin{center}
\includegraphics[width=0.32\textwidth]{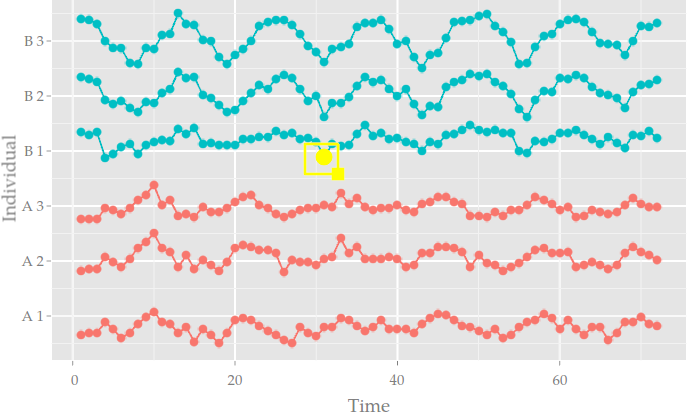}
\includegraphics[width=0.32\textwidth]{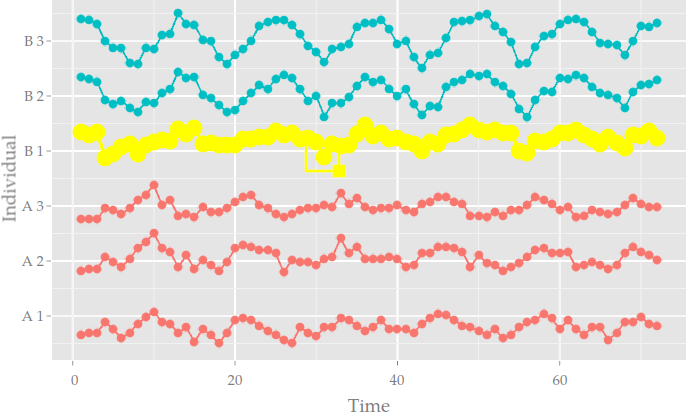}
\includegraphics[width=0.32\textwidth]{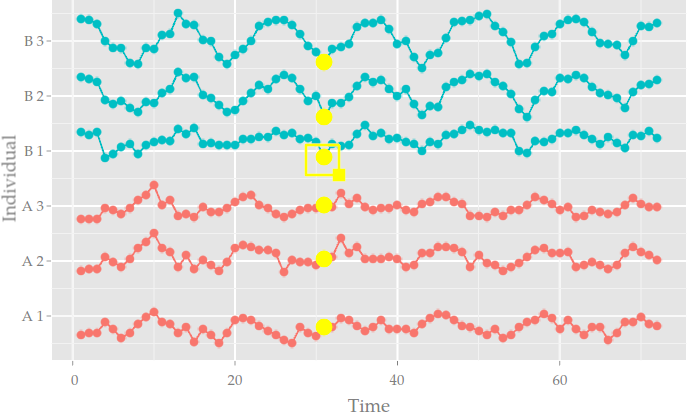}
\end{center}
\caption{\label{fig:self-linking}When a single point is brushed,
there could be three modes of highlighting: (left) only a single
point is highlighted; (center) all points for that series are
highlighted, by treating the line group indicators as the categorical
linking variables; (right) all points for that time are highlighted,
by using the time as a linking variable.}
\end{figure}

A more flexible format is provided by melting the data
into the ``long data'' format (Table \ref{tab:long-data}).
In this format it is easy to realize brushing and self-linking
in different ways: the user can select a single point, and
(1) only this point is highlighted, (2) all points for
that line (e.g. V1) are highlighted, or (3) all points
for that time are highlighted (Figure \ref{fig:self-linking}).
The latter two are achieved by treating the line group
indicator or ``Time'' as a categorical linking variable,
respectively.

\begin{table}[h]
\small
\begin{center}
\begin{tabular}{ccc}
\begin{tabular}{|c|ccc|ll|}
\hline 
\multicolumn{4}{|c|}{Variables} & \multicolumn{2}{c|}{Parameters}\tabularnewline
\hline 
Time & V1 & V2 & V3 & \texttt{.brushed} & \texttt{.color} \tabularnewline
\hline 
1 & 3.1 & 27 & 11.9 & FALSE & red \tabularnewline
2 & 3.4 & 23 & 12.5 & FALSE & blue \tabularnewline
$\vdots$ & $\vdots$ & $\vdots$ & $\vdots$ & $\vdots$ & $\vdots$ \tabularnewline
\hline 
\end{tabular}
& &
\begin{tabular}{|c|ccc|ll|}
\hline 
\multicolumn{4}{|c|}{Variables} & \multicolumn{2}{c|}{Parameters}\tabularnewline
\hline 
Time & V1 & V2 & V3 & \texttt{.brushed} & \texttt{.color} \tabularnewline
\hline 
1 & 3.1 & 27 & 11.9 & FALSE & red \tabularnewline
\red{2} & \red{3.4} & \red{23} & \red{12.5} & \red{TRUE} & \red{yellow} \tabularnewline
$\vdots$ & $\vdots$ & $\vdots$ & $\vdots$ & $\vdots$ & $\vdots$ \tabularnewline
\hline 
\end{tabular}
\end{tabular}
\end{center}
\normalsize
\caption{\label{tab:wide-data}Basic tabular form of data underlying plot (left), think of this as the ``wide data'' format. Each row contains values recorded at one time point. The aesthetic parameters are associated with one time point. Brushing on this form will highlight the points for a particular time (right), three points if all three series are drawn. It is probably more desirable for the behavior to be different: that selecting a single point will highlight all the values for that series, or only a single point, which can be achieved by a data re-structuring.}
\end{table}

\begin{table}[htp]
\begin{center}
\begin{tabular}{|c|cc|ll||ll||ll|}
\hline 
\multicolumn{1}{|c}{} & \multicolumn{2}{c|}{} & \multicolumn{2}{c||}{Single point}& \multicolumn{2}{c||}{All V1}& \multicolumn{2}{c|}{All Time$=2$}\tabularnewline
\hline 
Time & Variable & Value & \texttt{.brushed} & \texttt{.color} & \texttt{.brushed} & \texttt{.color} & \texttt{.brushed} & \texttt{.color}\tabularnewline
\hline 
1 & V1 & 3.1 & TRUE & red & \textcolor{red}{TRUE} & \textcolor{red}{yellow} & TRUE & red \tabularnewline
2 & V1 & 3.4 & \textcolor{red}{TRUE} & \textcolor{red}{yellow} & \textcolor{red}{TRUE} & \textcolor{red}{yellow} & \textcolor{red}{TRUE} & \textcolor{red}{yellow}\tabularnewline
1 & V2 & 27 & FALSE & red & FALSE & red & FALSE & red\tabularnewline
2 & V2 & 23 & FALSE & blue & FALSE & blue & \textcolor{red}{TRUE} & \textcolor{red}{yellow} \tabularnewline
1 & V3 & 11.9 & FALSE & red & FALSE & red & FALSE & red \tabularnewline
2 & V3 & 12.5 & FALSE & blue & FALSE & blue & \textcolor{red}{TRUE} & \textcolor{red}{yellow}\tabularnewline
$\vdots$ & $\vdots$ & $\vdots$ & $\vdots$ & $\vdots$ & $\vdots$ & $\vdots$ & $\vdots$ & $\vdots$ \tabularnewline
\hline 
\end{tabular}
\end{center}
\caption{\label{tab:long-data}The ``long data'' format, which is called the melted form in \citet{reshape}. This format allows a lot of flexibility. The ``Variable'' column can be used as a categorical linking variable, so that all points corresponding to ``V1'' are highlighted, when any one is selected, or a single point can be highlighted in the simplest brushing style, by changing the parameters of just that row. It would also be possible to use this form to use ``Time'' as a categorical variable for linking, and highlight all values recorded at a particular time. }
\end{table}

The long data format is typically the basic format
for longitudinal data, where there may be differing
number time points per subject, and measured at
different times. So this approach to implementing
the brushing works here, too.

\subsection{Linking between plots}

Linking between plots builds a reactive brushing chain that
when the data points on one plot are brushed, then they are
highlighted on all the plots. In the normal cases of
\texttt{\textbf{cranvas}}, data behind the plots is unique,
so the brush interaction will modify the parameter attached
to the data and trigger the listeners of plots to highlight
the corresponding part. However, linking between a time
series plot and other plots is different, because the
data to create the time plot is in the long data format,
while the data to create the other plots, like scatterplots
or histograms, are in the wide data format. Hence, a link
between the two data formats must be constructed.

\citet{xie2014reactive} delineated how to link two data
objects. First a linking variable must be pointed out, then two listeners
are attached on the two objects. If the \texttt{.brushed} parameter
switches in one dataset, then the listener is triggered. And if any
observations from the second dataset have the same value in the linking
variable as the first dataset, then the corresponding \texttt{.brushed}
parameter in the second dataset will be changed. 

Note that the linking between wide data and long data is not a one-to-one
linking. ``Time'' is the linking variable between two formats.
In the direction from wide to long data, each entry in the
wide data can project to multiple entries in the long data. In the
opposite direction, an entry in the long data will map to one entry in
the wide data. The unbalanced linking could produce a problem, as
shown in Table \ref{tab:wide-long-linking}.

This problem can be solved by cutting off the backward linking.
To facilitate the cutoff, two signals are added respectively
in the listeners of the two data objects. When one listener
is triggered, the signal will be turned on until the listener
finishes its work. During this period, the other listener cannot work.
As in Table \ref{tab:wide-long-linking}, the arrow from  (b) to  (c)
will be cut off.

\begin{center}
\begin{figure}[htp]
\begin{centering}
\begin{tabular}{cl}
\multirow{5}{*}[2.1in]{\includegraphics[width=0.45\textwidth]{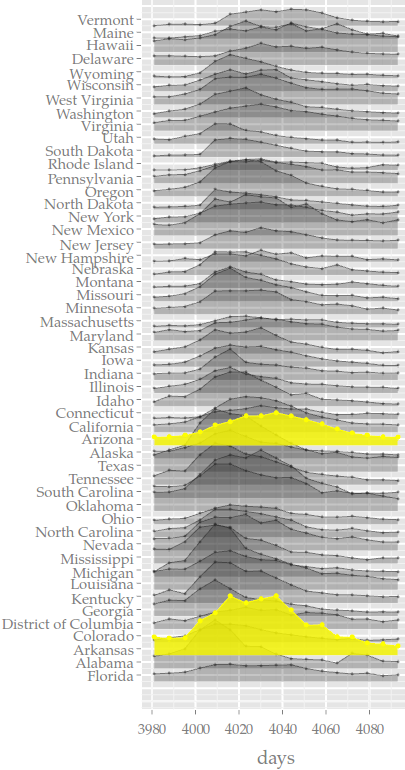}}
& \includegraphics[width=0.48\textwidth]{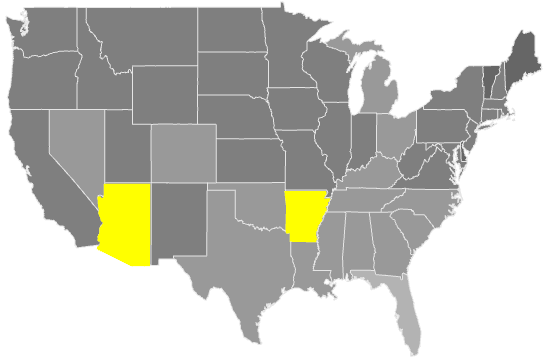} \tabularnewline
& \tabularnewline
& \tabularnewline
& \includegraphics[width=0.4\textwidth]{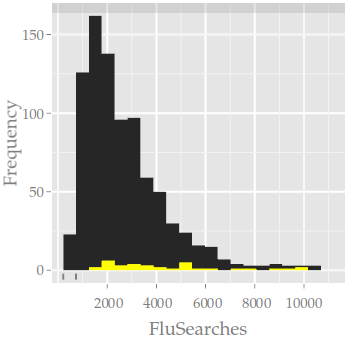} \tabularnewline
& \tabularnewline
\end{tabular}
\end{centering}
\caption{\label{fig:linking-plots}Linking between different plots
for the Google Flu Trends data from November 24, 2013 -- March 16, 2014.
(Left) Time series faceted by state (see \url{https://vimeo.com/112528131}
for the transition to the facets), (top right) choropleth map grey
scale indicating time of the peak in the series, light is earlier,
(bottom right) histogram of the number of searches. Two states in
the map are brushed, which highlights all flu searches from these
states in the time series and histogram.}
\end{figure}
\end{center}

Figure \ref{fig:linking-plots} shows the linking between
a longitudinal time plot, a map, and a histogram. The data
come from the Google Flu Trends
(\url{http://www.google.org/flutrends/}).

\begin{table}[htp]
\begin{center}
\begin{tabular}{c|cc|cc}
\multicolumn{5}{c}{(a) Long data format, used in time plots}\tabularnewline
\hline 
Time & Variable & Value & \texttt{.brushed} & \texttt{.color} \tabularnewline
\hline 
$\vdots$ & $\vdots$ & $\vdots$ & $\vdots$ & $\vdots$ \tabularnewline
\textcolor{red}{2} & \textcolor{red}{V1} & \textcolor{red}{3.4} & \textcolor{red}{TRUE} & \textcolor{red}{blue} \tabularnewline
2 & V2 & 23 & FALSE & blue \tabularnewline
2 & V3 & 12.5 & FALSE & blue \tabularnewline
$\vdots$ & $\vdots$ & $\vdots$ & $\vdots$ & $\vdots$ \tabularnewline
\hline 
\end{tabular}

$\Downarrow$

\begin{tabular}{c|ccc|cc}
\multicolumn{6}{c}{(b) Wide data format, used in other plots}\tabularnewline
\hline 
Time & V1 & V2 & V3 & \texttt{.brushed} & \texttt{.color} \tabularnewline
\hline 
$\vdots$ & $\vdots$ & $\vdots$ & $\vdots$ & $\vdots$ & $\vdots$ \tabularnewline
\textcolor{red}{2} & \textcolor{red}{3.4} & \textcolor{red}{23} & \textcolor{red}{12.5} & \textcolor{red}{TRUE} & \textcolor{red}{blue} \tabularnewline
$\vdots$ & $\vdots$ & $\vdots$ & $\vdots$ & $\vdots$ & $\vdots$ \tabularnewline
\hline 
\end{tabular}

$\Downarrow$

\begin{tabular}{c|cc|cc}
\multicolumn{5}{c}{(c) Long data format, used in time plots}\tabularnewline
\hline 
Time & Variable & Value & \texttt{.brushed} & \texttt{.color} \tabularnewline
\hline 
$\vdots$ & $\vdots$ & $\vdots$ & $\vdots$ & $\vdots$ \tabularnewline
\textcolor{red}{2} & \textcolor{red}{V1} & \textcolor{red}{3.4} & \textcolor{red}{TRUE} & \textcolor{red}{blue} \tabularnewline
\textcolor{red}{2} & \textcolor{red}{V2} & \textcolor{red}{23} & \textcolor{red}{TRUE} & \textcolor{red}{blue} \tabularnewline
\textcolor{red}{2} & \textcolor{red}{V3} & \textcolor{red}{12.5} & \textcolor{red}{TRUE} & \textcolor{red}{blue} \tabularnewline
$\vdots$ & $\vdots$ & $\vdots$ & $\vdots$ & $\vdots$ \tabularnewline
\hline 
\end{tabular}
\end{center}
\caption{\label{tab:wide-long-linking}Linking between the long and wide data
format may produce a problem. Suppose we start from brushing a point  (V1
at time 2) in the long data  (a). Then the listener of the long data
is triggered and changes the .brushed parameter for time 2 in the
wide data  (b). Then the listener of the wide data is triggered and
switches the .brushed parameter for all the observations at time 2
in the long data  (c), which will update (a) but make a conflict.}
\end{table}

\subsection{Additional linking issues\label{sub:Linking-of-the-addition}}

Besides the issue of wide data and long data, there are other datasets
created during the interactions that could produce some linking issues,
such as the follows.
\begin{itemize} \itemsep 0in
\item The polygon data from the area layer. 
The area layer does not
only need the values in the time series, but also the baseline to
form many polygons. To draw the polygons, the data must be rearranged
in an order of polygon vertexes. Each polygon is made of four vertexes:
two from the time series and two from the baseline. Because the polygon
layer should listen to the point and line layers, the link between
the original data and the polygon data is one-way.

\item Copies of the dataset during the incremental operation.
When adopting the incremental procedure in Section 
\ref{sub:Two-procedures}, the variations of original dataset will
be created with multiple stages of the interactions. However, these
additional datasets are not required to get linked, because we only
make the copies of the coordinates, not the properties. To use the
coordinates, we combine them with the properties in the last minute,
so there will not be any linking issues.

\item Additional areas from the vertical faceting.
The example of vertical faceting is shown in Figure \ref{fig:y-wrapping}
and Figure \ref{fig:Additional-data} explained how the interaction
creates the additional areas. In Figure \ref{fig:Additional-data},
the black dots are given by the time series, and the red dots are
created during the cropping step of faceting, by the choice of
cutting lines. The black and red dots are mixed in some order to
form the shaded polygons. Whenever a point is brushed, one, two,
or even more cropped polygons should be highlighted. Hence the
two-direction linking between the cropped polygons and points
should be constructed. Note that this is a one-to-$n$ mapping,
and the linking variable is the point ID, which should be assigned
to the polygons when the red dots are generated.

\begin{center}
\begin{figure}[htp]
\begin{centering}
\includegraphics[width=0.48\textwidth]{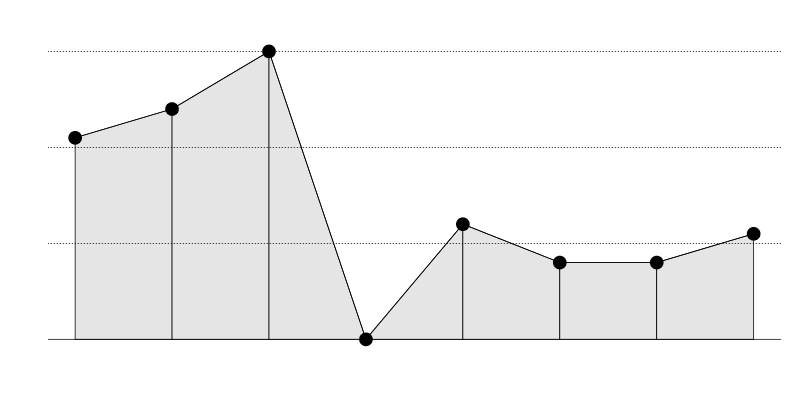}
\includegraphics[width=0.48\textwidth]{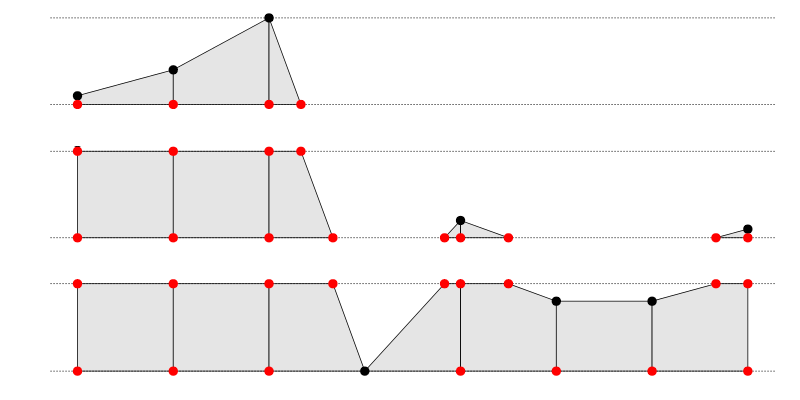}
\end{centering}
\caption{\label{fig:Additional-data}Additional data created from the vertical
faceting. The black dots are given by the time series, and the red
dots are created during the cropping step of faceting.}
\end{figure}
\end{center}

\end{itemize}

\section{Querying}

In each of the examples, and plots shown the time axis is
simply drawn using consecutive integers. This is necessary
for convenience and generalizability. To know which actual
time value requires querying, by mousing over the display,
if labels have been set up in detail, the user can learn what
day, month, year or individual identifier is under the cursor.

\section{Conclusions and future work}

This paper describes how interactions can be added to temporal displays by using sequences of affine transformations. This approach proves a rich variety of ways to slice and dice temporal data to explore seasonality, dependence, trends, anomalies and individual differences. These interactive temporal displays can be embedded in a large interactive graphics software system, enabling linking between plots to explore more general data where temporal components are just one aspect of many. Not everything is solved in terms of additivity, speed and interaction direction, but most common actions are possible for reasonably sized data. 

Optimal aspect ratio for temporal displays is an unsolved problem,
and is something that we have grappled with in the implementation.
According to \citep{cleveland1993} time series should be ``banked''
to 45$^o$, which means that on average the angle of the lines, in
comparison to the $x$- or $y$-axes is 45$^o$. This is computed and
used in the initial plots in \texttt{\textbf{cranvas}}, but as
wrapping and faceting are conducted it probably should be
re-calculated. Ideal integration of aspect ratio re-drawing with
plot interactions could be examined in new work.

Future work would extend the interactions to include transforming between euclidean and polar coordinates, and between time and frequency domains. Earlier work in Dataviewer \citep{BAHM88} allowed users to interactively lag time series to generate lag plots to explore auto-correlation. This could be reasonably be accomplished similarly to the interactions described here.

Using data transformations to generate interactions is efficient but it assumes the components are ordinal. This is not necessarily true, for example, in the flu searches series (Figure \ref{fig:linking-plots}) were faceted on the categorical variable state. This requires that states are first recoded to numerical values. In R, this is implicitly alphabetical order. However, because the implementation is created in R, the factor levels can be re-ordered easily, enabling the recoding to numerical value to be quite fluid. In the flu searches example, the states were re-ordered by earliest peak of searches.

This paper focused on interactive graphics for multivariate time series and longitudinal data, but the ideas should extend some to other temporal-context data. 

\section*{Acknowledgements}
This work was started with funding from Google Summer of Code,
then partially supported by an unrestricted fellowship from
Novartis, and by National Science Research grant DMS0706949.

\bibliographystyle{natbib}
\bibliography{cheng2014interactivity}

\end{document}